\documentclass[%
superscriptaddress,
showpacs,
 amsmath,amssymb,
 aps,
prb,
floatfix,
twocolumn
]{revtex4-1}

\usepackage{lipsum}

\usepackage{graphicx}
\usepackage{dcolumn}
\usepackage{bm}
\usepackage{MnSymbol}
\usepackage{color}
\usepackage{dsfont}
\usepackage{hyperref}
\usepackage{float}

\usepackage{color}
\usepackage{ulem}
\usepackage{cancel}
\definecolor{violet}{rgb}{0.58, 0.0, 0.83}
\newcommand{\mkadd}[1]{{\color{violet}{#1}}}

\begin{document}

\global\long\def\bra{\langle}
\global\long\def\brat#1{\langle#1|}
\global\long\def\ket{\rangle}
\global\long\def\kett#1{|#1\rangle}
\global\long\def\d{\partial}
\global\long\def\s#1{\mathcal{#1}}
\global\long\def\ex#1{\bra#1\ket}
\global\long\def\p#1{\left(#1\right)}

\global\long\def\qb#1#2{\langle#1|#2\rangle}
\global\long\def\mel#1#2#3{\langle#1|#2|#3\rangle}

\global\long\def\ple#1{\left(#1\right.}
\global\long\def\pre#1{\left.#1\right)}

\global\long\def\dag{\dagger}


\title{Absence of Thermalization in Finite Isolated Interacting Floquet Systems}
\author{Karthik Seetharam}
\email{kseethar@caltech.edu}
\affiliation{Institute for Quantum Information and Matter, Caltech, Pasadena, California 91125, USA}
\author{Paraj Titum}%
\affiliation{Institute for Quantum Information and Matter, Caltech, Pasadena, California 91125, USA}
\affiliation{Joint Quantum Institute and Joint Center for Quantum Information and Computer Science,
NIST/University of Maryland, College Park, Maryland 20742, USA}
 \email{paraj@umd.edu}
 \author{Michael Kolodrubetz}%
\affiliation{Materials Sciences Division, Lawrence Berkeley National Laboratory, Berkeley, California 94720, USA}
\affiliation{Department of Physics, University of California, Berkeley, California 94720, USA}
\affiliation{Department of Physics, University of Texas at Dallas, Richardson, Texas 75080, USA}
 \email{mkolodru@utdallas.edu}
\author{Gil Refael}
\email{refael@caltech.edu}
\affiliation{Institute for Quantum Information and Matter, Caltech, Pasadena, California 91125, USA}

\date{\today}

\begin{abstract}
Conventional wisdom suggests that the long time behavior of isolated interacting periodically driven (Floquet) systems is a featureless maximal entropy state characterized by an infinite temperature. Efforts to thwart this uninteresting fixed point include adding sufficient disorder to realize a Floquet many-body localized phase or working in a narrow region of drive frequencies to achieve glassy non-thermal behavior at long time. Here we show that in clean systems the Floquet eigenstates can exhibit non-thermal behavior due to finite system size. We consider a one-dimensional system of spinless fermions with nearest-neighbor interactions where the interaction term is driven. Interestingly, even with no static component of the interaction, the quasienergy spectrum contains gaps and a significant fraction of the Floquet eigenstates, at all quasienergies, have non-thermal average doublon densities. We show that this non-thermal behavior arises due to emergent integrability at large interaction strength and discuss how the integrability breaks down with power-law dependence on system size.

\end{abstract}

\maketitle

\section{\label{sec:Intro} Introduction}

Periodically driven systems offer the tantalizing potential to engineer and control the collective behavior of quantum systems, which has been extremely useful in realizing novel phases of matter\cite{Oka2009,KBRD,Lindner2011,Jiang2011,Kitagawa2011}. Often these driven systems support phases without any equilibrium analog such as time crystals and the so-called anomalous Floquet topological phases \cite{KhemaniPRL2016,Else_bauer_Nayak_PRL2016,Yao_2016,KeyserlingkSondhi2016,Keyserlingk_PRB2016-II,Rudner2013,TitumPRX2016,Po_Vishwanath2017,KeyserlingkSondhi2016,ElseNayak2016,PotterVishwanath2016,RoyHarper2016,PoVishwanath2016,HarperRoy2017,PotterMorimoto2017,RoyHarper2017,PoPotter2017,PotterFidkowski2017}. Recently, novel Floquet phases have been observed experimentally in a variety of systems such as trapped ions, cold atoms, NV centers, and photonic devices\cite{Monroe_2017,Lukin_2017,Jotzu2014,Rechtsman2013,Bloch_Nat_Phys2017,AidelsburgerBloch2013,MiyakeKetterle2013,ParkerChin2013}. The high degree of control in these artificially engineered systems allows for precise implementation of periodically driven Hamiltonians and for easy measurements of local observables.

Predicting the long time dynamics of isolated interacting quantum systems remains a challenge. The generic behavior of such systems may be classified as thermal or non-thermal. The  behavior of time-independent thermal Hamiltonians is well-described by the eigenstate thermalization hypothesis (ETH) \cite{Nandkishore_Review_2015,Alessio_review_2016,Deutsch1991,Srednicki1994,RigolOlshanii2008}. According to the ETH, at long times and in the thermodynamic limit, all local observables asymptotically reach a value as given by a thermal density matrix with a temperature corresponding to the energy density of the initial state. An analogous claim can be made for periodically driven systems for which understanding such thermalization is not only crucial for experimental efforts, but also for realizing uniquely nonequilibrium phases. Given that energy is not conserved in such systems, the long-time thermal state is characterized by infinite temperature and maximal entropy\cite{Alessio_Rigol_2014_PRX,Lazarides_Moessner_2014_PRL,Lazarides_Moessner_2014_PRE}. This means that, in addition to being thermal, the long time dynamics of isolated interacting periodically driven systems is independent of the choice of the initial state. 

\begin{figure}
\includegraphics[width=\columnwidth]{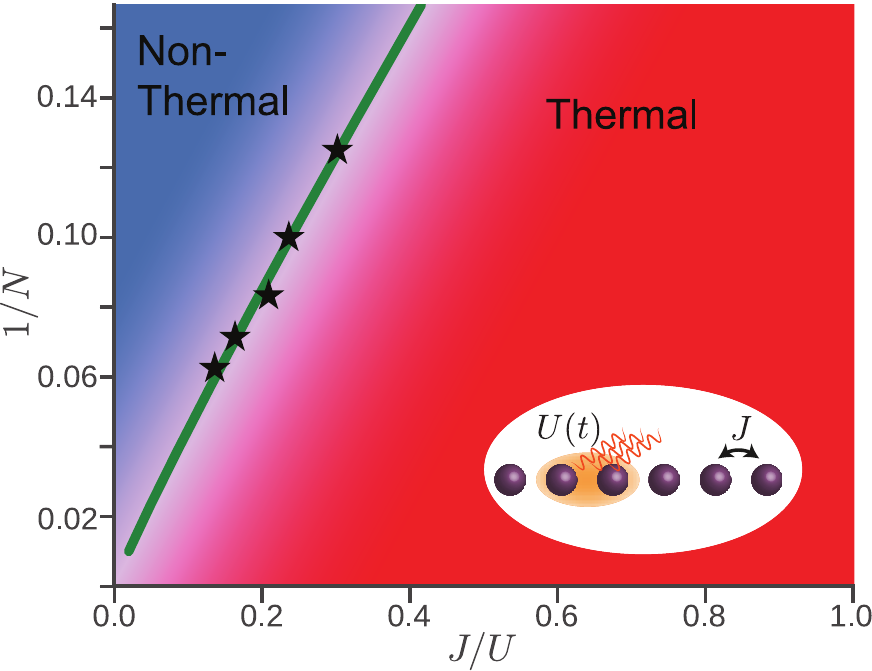}

\caption{Phase diagram showing the thermal (red) and non-thermal (blue) behavior of the periodically driven model described in Eq. \ref{eq:hamiltonian}. We see that at finite size, $N$, and large $U/J\gg 1$, the periodically driven chain exhibits non-thermal behavior. In the thermodynamic limit, this region vanishes. The fitting points (black stars) indicate the approximate crossover region as obtained from exact diagonalization. The crossover line (green) between the thermal and non-thermal region is a power-law fit to the black stars $(\frac{J}{U})_\mathrm{c}\approx 2.9 N^{-1.1}$.}
\label{fig:phasediagram}
\end{figure} 

Such a featureless state is uninteresting from both a theoretical and experimental point of view, so the questions remain as to when and how non-infinite-temperature behavior can be achieved and controlled. Of course, if one opens the system to engineered dissipation, it is possible to induce quasithermal steady states with finite temperature and chemical potential \cite{seetharam2015,IadecolaChamon2015-1,IadecolaChamon2015-2,DehghaniMitra2014}. However, for those systems which are well-isolated (e.g. cold atoms), alternative routes to non-thermal behavior are currently being explored. One example is quantum integrable models, which have an extensive number of local conserved quantities such that the long-time dynamics of observables are, in many cases, characterized by a generalized Gibbs ensemble (GGE)\cite{Vidmar_Rigol_GGE_Review,PolkovnikovColloquium11}. Recently, it has been shown that with the appropriate choice of disorder, there is an emergent notion of integrability associated with many-body localization (MBL)\cite{Nandkishore_Review_2015,Abanin_Papic_MBLReview_2017}. Furthermore, there is some evidence for partial breakdown of thermalization in translationally-invariant models\cite{GarrisonFisher2017,YaoMoore2017,SmithMoessner2017}. These ideas of localization and integrability have also been extended to periodically driven systems\cite{Alessio_Polkovnikov_annals_2013,Ponte2014,Ponte2014b,Lazarides_Das_Moessner_PRL_2015,KhemaniPRL2016,AgarwalaSen2017}. Finally, even if a system eventually thermalizes to the infinite-temperature state, it is possible that the time scale to approach such a state is quite long, and thus there exists a ``prethermal" regime where interesting physics can be explored \cite{Weidinger_Knap_2017,MachadoArxiv2017,BukovDemler2015,ElseNayak2017PRX,ZengSheng2017,AbaninHuveneers2017,KuwaharaSaito2016}.

In this paper, we explore the effects of strong driving on an integrable model, the spinless fermionic Hubbard model with nearest-neighbor interactions. The undriven model is exactly solvable via Bethe ansatz \cite{SirkerXXZ}. We show that, in the thermodynamic limit, the introduction of driving leads to uncontrolled heating for all finite interaction strengths. Remarkably, at finite size, it is possible to recover non-thermal behavior for a large region in the parameter space of interactions, both for very weak and strong interactions. In both of these limits, the non-thermality is governed by a nearby (in interaction strength) integrable point that controls the behavior at finite size, a notion we will term nearly-integrable. We show that above a certain interaction scale determined by the system size, the system crosses over from non-thermal to thermal. Similar ideas about the existence of non-thermal states at finite size have been discussed in the equilibrium context in Ref.~\onlinecite{RigolSantos2010PRA,SantosRigol2010PRE}.
We also note that recent work on integrability breaking in Floquet systems has focused on the high-frequency limit \cite{ClaeysCaux2017}, studying the onset of heating as frequency is lowered. By contrast, in this work we analyze finite size scaling in regimes of highly resonant interactions as a function on interaction strength and discover nearly-integrable behavior. The complementary results provide a potential finite size scaling foundation upon which to build an analytical theory of integrability-breaking and the breakdown of the high freuqency expansion in Floquet systems.
  
This paper is organized as follows. In section \ref{sec:model}, we introduce the model, provide a rudimentary overview of Floquet theory, describe the properties of the undriven model with a particular emphasis on finite size, and finally provide an intuitive discussion of thermalization in periodically driven systems. In section \ref{sec:Modulated Interaction}, we present the basic data of the driven model including spectral information, doublon density, and time evolution of a few representative initial states, all as a function of the interaction strength. In section \ref{sec:Scaling}, we discuss the results of finite size scaling that distinguish the non-thermal and thermal regions. In section \ref{sec:Integrability and its breakdown}, we show that the origin of the non-thermal region is due to integrability and that its subsequent breakdown is responsible for the crossover to the thermal region. Finally, in section \ref{sec:Discussion}, we recapitulate the results and discuss future directions of research.        

In the appendices, we further establish the robustness of our results to changes in the model. First, in appendix \ref{sec:Frequency Dependence}, we show that these results are universal in the highly resonant (i.e., low frequency) regime where $\Omega/J\sim 1$. Furthermore, we show how, at intermediate frequencies, the precise structure of the rare resonances dominates the behavior of the spectral variance of the doublon density. At sufficiently high frequencies, i.e., those above the many-body bandwidth, we recover the usual result of high frequency expansions that the dynamics are given by the time-averaged Hamiltonian, which, in our case, is just that of a free fermion static metal. In appendix \ref{sec:Waveform Dependence}, we show that the non-thermal regime exists for other waveforms. Specifically, we show that as we interpolate from a square wave to a single harmonic, the non-thermal regime exists albeit weakened by a larger crossover region. Therefore, we conclude that the non-thermal region is robust, suggesting that the general concept of near-integrability persisting at finite size occurs independent of the exact details of the model. \footnote{Although, of course, the precise scaling and crossover behavior indeed should depend on model.}

\section{\label{sec:model} Model}

In this section, we first introduce a one-dimensional model for a closed periodically-driven system of spinless interacting fermions. We discuss the Floquet states which form a convenient time-dependent basis for study of a time-periodic Hamiltonian. Next, we provide some intuition for the behavior of the undriven model. Finally, we review some known results on thermalization in closed Floquet systems.

\subsection{\label{subsec:Hamiltonian} Hamiltonian}

Consider a Hamiltonian of spinless fermions interacting via nearest-neighbor Hubbard interactions,
\begin{eqnarray}
H&=&J\sum_{i}(c_{i}^{\dag}c_{i+1}+c_{i+1}^{\dag}c_{i})+U(t)\sum_{i}n_{i}n_{i+1}\label{eq:hamiltonian}\nonumber\\
\textrm{with,} &&U(t)=U_{0}f_{U}(t)
\end{eqnarray}
where $n_i=c^\dag_i c_i$ is the fermion density and $U(t)$ is the time-periodic nearest-neighbor interaction coupling (see inset of Figure \ref{fig:phasediagram}). Different driving protocols with angular frequency $\Omega$ are set by $f_U(t)=f_U (t+\frac{2\pi}{\Omega})$. Throughout this work, we consider the case of the lattice at half-filling and driving protocols with no static component $\int_0^{T} f_U(t) dt = 0$. 

\subsection{\label{subsec:Floquet Theory} Floquet Theory}

For a time-periodic Hamiltonian $H(t+T)=H(t)$, the Floquet theorem states that one may always decompose the time evolution operator as $U(t,t_0)=P(t,t_0)e^{-iH_F[t_0](t-t_0)}$ where $H_F[t_0]$ is a time-independent operator known as the stroboscopic Floquet Hamiltonian and $P(t,t_0)$, commonly called the micro-motion operator, is periodic in both arguments. The latter governs the ``fast'' intra-period evolution whereas the former governs the ``slow" stroboscopic dynamics. Here $t_0$ is the choice of initial time for the evolution, which is equivalent to the choice of initial phase of the drive. Throughout this manuscript we use the Floquet gauge choice $t_0=0$ and drop the argument $t_0=0$ for convenience. More discussion of gauge choices can be found in appendix \ref{sec:Expansion}. 

To obtain Floquet quasienergies, $\mathcal{E}$, and eigenstates, $|n_F\rangle$, throughout this manuscript we proceed by constructing $U(T,0)$ explicitly and diagonalizing $H_{F}=\frac{i}{T}\mathrm{log}U(T,0)$. This method is useful for periodic drives where $U(T,0)$ can be easily written as a product of a few evolution operators, such as a square wave.

\subsection{\label{subsec:Undriven Model} Undriven Model}
The undriven model is integrable as it is equivalent to an XXZ chain via Jordan-Wigner transformation. In this case, one may compute the spectrum in the thermodynamic limit using Bethe ansatz. Let us, however, obtain some intuition for the simple limits of the undriven model while explicitly keeping track of finite size. For the case of pure nearest-neighbor hopping, the many-body bandwidth for a system of $M$ fermions in $N>M$ sites is $\leq4MJ$, which, at any fixed density, scales as $NJ$. For the case of pure interaction, where for the moment we assume a nonzero static $U_0$, the many-body bandwidth is $U_0(M-1)$, which, at any fixed density, scales as $NU_0$. Note the factor of $(M-1)$ is the maximum number of doublons, defined as $\bar{n}_i=n_i n_{i+1}$, one can obtain for a finite chain system without periodic boundary conditions. With both hopping and interactions, in the case where $U_0 > NJ$, the doublon spacing $U_0$ is bigger than the bandwidth induced by hybridization, via hopping, of the doublon sectors. Hence, the doublon sectors disperse in energy but still are separated from each other. In the thermodynamic limit ($N\rightarrow\infty$) for any finite $U_0$, the doublon sectors, from a spectral point of view, merge together. The intuition gleaned from this spectral analysis is that for sufficiently large interaction $U_0$ at a given finite size, doublon character seems to persist in the eigenstates, i.e., doublons are almost conserved. This finite size persistence is a simple example of what we term as near-integrability. Indeed, in this particular case, since the undriven model is Bethe ansatz integrable \cite{SirkerXXZ}, there is always an extensive set of conserved quantities, which, at infinite $U_0$, will again conserve doublons.  However, as we will show in this work, the near-integrability behavior in the presence of strong drive is significantly different and more subtle.

\subsection{\label{subsec:Interactions in Closed Floquet Systems} Thermalization in Floquet Systems}

Before delving into details of finite size scaling in our specific model, let us first review the generic expectations about thermalization and the role of interactions in closed systems. An undriven ``thermal'' system is often defined as that satisfying the eigenstate thermalization hypothesis (ETH). According to the ETH, eigenstates with similar energy will yield similar expectation values of local observables. Therefore, for an arbitrary initial state with small energy fluctuations, measurement of a local observable at late time may be replaced with measurement in the microcanonical ensemble at the same energy. As in conventional statistical mechanics, fluctuations of macroscopic conserved quantities vanish in the thermodynamic limit, leading to equivalence of ensembles.

Unlike static Hamiltonians, the presence of periodic driving destroys energy conservation and hence the ``microcanonical" state is now spread over all energies; such a uniform state with no constraints is just an infinite temperature Gibbs state. Therefore, the long time steady state of a generic periodically driven interacting system is intuitively expected to be the infinite temperature diagonal ensemble \cite{Bukov_Polkovnikov_2016_PRB,Alessio_Rigol_2014_PRX,Lazarides_Moessner_2014_PRE,Lazarides_Moessner_2014_PRL}. This means that the expectation value of a time-averaged local observable, $\mathcal{O}(t)$, starting from an arbitrary initial state, $\kett{\psi_0}$, is

\begin{eqnarray}
\overline{\ex{\s O(t)}} & = & \mathrm{lim}_{\tau\rightarrow\infty}\frac{1}{\tau}\int_{0}^{\tau}dt\mel{\psi_{0}}{U^{\dag}(t)\s OU(t)}{\psi_{0}}\nonumber\\
 & = & Tr[\rho_{\infty}\s O]\mkadd{,}
\end{eqnarray}
where $U(t)$ is the time evolution operator and $\rho_{\infty} = \mathrm{Dim}[\s H]^{-1}\mathbb{I}$ with $\mathrm{Dim}[\mathcal{H}]$ denoting the dimension of the Hilbert space. An important consequence of such an ensemble is that the long-time-averaged steady state value of $\mathcal{O}$ is independent of the initial starting state. 

All of these arguments about ETH and Floquet-ETH (the term we will use to characterize the infinite temperature ensemble) rely on generic and mostly unconstrained mixing of states via evolution under the Hamiltonian. This is the quantum analog of dynamical chaos leading to ergodicity in classical dynamical systems. Classically, an integrable system has an extensive number of mutually conserved quantities that destroy ergodicity; hence such systems certainly do not satisfy equilibrium statistical mechanics. In the quantum mechanical scenario, we will refer to integrability as a system with an extensive number of mutually commuting locally (additive) conserved quantities. The intuition here is the same as the classical case - the evolution of the states is highly constrained and so mixing does not really occur. With this understanding, it is clear that integrability yields non-thermal behavior.

\section{\label{sec:Modulated Interaction} Modulated Interaction}

We now return to the driven case where the resonant interaction has no static value and is modulated with angular frequency $\Omega$. Unless otherwise noted, we restrict ourselves to the case where the driving frequency is much smaller than the many-body bandwidth and to a square wave drive
\begin{eqnarray}
f_{U}(t) & = & \begin{cases}
1, & 0\leq t<T/2,\\
-1, & T/2\leq t<T,
\end{cases}
\end{eqnarray}
where $T=2\pi/\Omega$ is the period. Building upon the generic intuition developed in section \ref{subsec:Interactions in Closed Floquet Systems}, we expect that in the regime of small frequency, the periodic driving will induce a large number of resonances which allow the system to explore the full Hilbert space and result in an infinite temperature ensemble. However, as we shall show in the following, this expectation gets modified at finite size and large driving amplitude $U/J\gg 1$ where from now on we drop the subscript on $U_0$, writing $U$ for brevity.

\begin{figure}
\includegraphics[width = 3.5in]{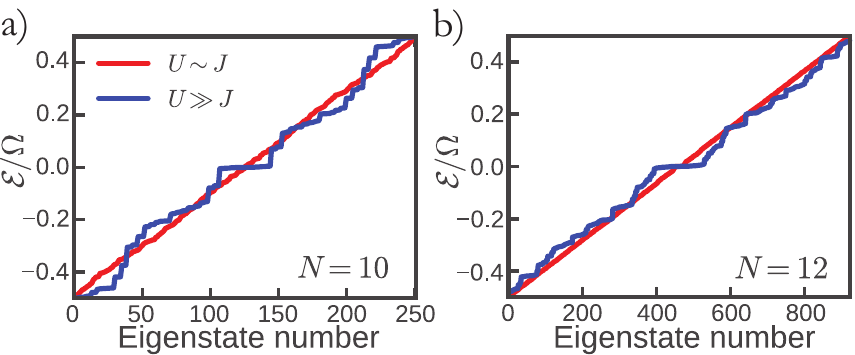}
\caption{Quasienergy spectrum for $N=10$ (a) and $N=12$ (b) at $\Omega/J=0.83$. Blue dots denote strong interaction $U/J=100$ and red dots denote weak interactions at $U/J=0.59$. We see that weak interactions give rise to a continuous spectrum. In contrast, the strong interactions yield separation of the spectrum into quasienergy plateaus reflecting the influence of doublons ($\sum_i n_in_{i+1}$). Increasing system size softens the plateaus.}
\label{fig:spectrum}
\end{figure}

For the special case of $J=0$, the model is trivially solvable, as any state picks up exactly the opposite phase during the first half of the cycle as during the second half, resulting in a perfect echo with $\mathcal E = 0$ for all eigenstates. However, in the presence of any small but finite $J$, the $U\rightarrow\infty$ limit is actually markedly different from $J=0$, as the perfect many-body echo is immediately destroyed. To gain simple intuition, we numerically solve for the quasienergy spectrum for $N=10$ and $N=12$ at half-filling.
The results for both system sizes, for two limiting cases --  $U \gg  J$ (blue) and $U \sim J$ (red) -- are shown in Figure \ref{fig:spectrum}. We have set $\Omega/J=0.83$ which is well below the many-body bandwidth, implying we are in the highly resonant regime. Remarkably, we see that the Floquet spectrum with large driven interaction ($U/J=100$ fixed for both sizes) shows plateau structures, which suggest that the influence of doublons is strong even when no static interaction is present. In contrast, for small interaction ($U/J=0.59$ fixed for both sizes), the Floquet spectrum looks continuous throughout the Floquet zone. We further note that increasing the system size while keeping the interaction fixed leads to a softening of the plateaus, suggesting that these effects may be related to the fact that our system is not in the thermodynamic limit. 

\subsection{\label{subsec:Doublon Correlations U} Doublon Density}

To systematically explore the presence of quasienergy plateaus in the Floquet spectrum, we calculate the density of doublons in each of the Floquet states.
\begin{equation}
\hat D=\frac{1}{\frac{N}{2}-1}\sum_{i}n_{i}n_{i+1}
\label{eq:doublon}
\end{equation}
The factor $N/2 -1$ in the denominator is the maximum number of doublons achievable for a chain of length $N$ at half-filling. This normalization factor ensures that the observable is bounded, $D \equiv \langle \hat D \rangle \in[0,1]$, and is independent of system size.

As discussed in Section \ref{subsec:Interactions in Closed Floquet Systems}, periodic driving is expected to lead to an infinite temperature ensemble, and as a result, any local observable measured in any Floquet state must yield the same value. In the infinite temperature ensemble at half-filling, one may explicitly calculate the expectation value of the doublon density 
\begin{eqnarray}
D &=&{{N}\choose{\frac{N}{2}}}^{-1}\mathrm{Tr}(\hat D)\nonumber\\
&=&\frac{1}{{{N}\choose{\frac{N}{2}}}\left(\frac{N}{2}-1\right)}\sum_{k=1}^{\frac{N}{2}-1}k{{\frac{N}{2}+1}\choose{k+1}}{{\frac{N}{2}-1}\choose{k}}=\frac{1}{2}
\label{eq:DoublonInf}
\end{eqnarray}
where ${n}\choose{k}$ denotes the binomial factor. Intuitively, one may understand this result as summing over $N/2$ particles with each particle having a neighbor with probability $1/2$ since the infinite temperature density matrix encodes no correlations. Hence, if we observe $D\ne0.5$ for a Floquet state, we may conclude that the state is by definition non-thermal. It is important to note that even if a state yields $D=0.5$, it is possible that another observable exists that can be measured which results in a value different from that given by an infinite temperature state. However, since our efforts to understand thermalization in this work focus on large $U$, we will use this observable as an indicator of non-thermality.

We examine the distribution of the doublon correlations by defining the variance of $D_n \equiv \langle n_F | \hat D | n_F \rangle$ over the Floquet eigenstates $|n_F\rangle$ as $\Sigma=\mathrm{var}_n(D_n)$. As we will see shortly, this spectral doublon variance will be quite useful in characterizing how the the entire spectrum changes as a function of coupling and system size.

\subsection{\label{subsec:Time Evolution} Time Evolution}

\begin{figure}
\includegraphics[width = 3.5in]{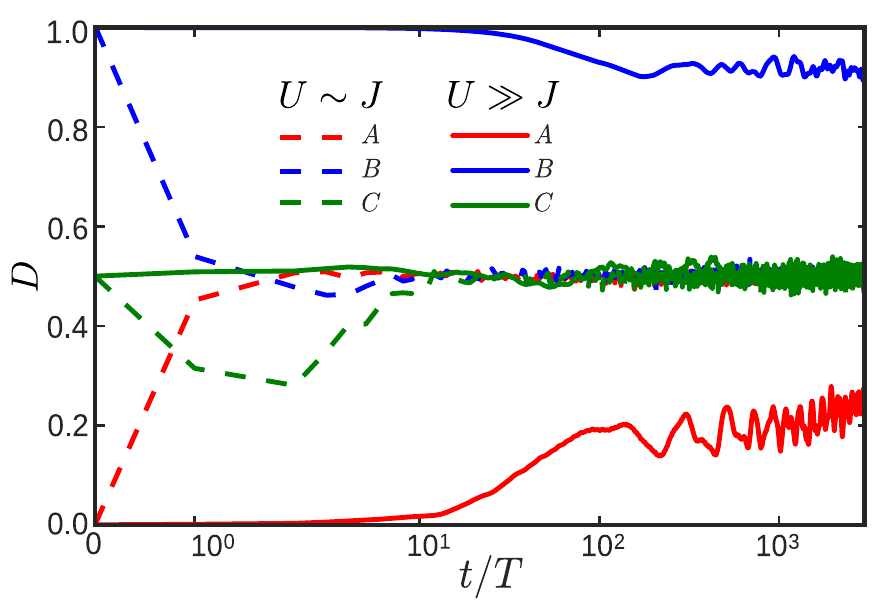}
\caption{Time evolution of three initial states for weak and strong interactions ($U/J=0.59$ and $U/J=100$ respectively): $A=\kett{101010...}$, $B=\kett{111...000}$, and $C=(C_{N/2}^{N})^{-1/2}\sum_{i=1}^{C_{N/2}^{N}}\kett i$. For weak interactions all states thermalize as expected. For strong interactions, the initial states with non-thermal doublon values ($A,B$) maintain non-thermal values over time whereas $C$ remains thermal.}
\label{fig:timeEvo}
\end{figure}

Let us now focus on the time-dependence of the doublon density for initial states which are not Floquet eigenstates and hence not stationary. We consider three example states: $A=\kett{101010...}$, $B=\kett{111...000}$, and $C=(C_{N/2}^{N})^{-1/2}\sum_{i=1}^{C_{N/2}^{N}}\kett i$ which are, respectively, a no-doublon state, a maximum-doublon state, and a state composed of an even superposition of all real space occupation basis states ($C_{N/2}^{N}={{N}\choose{N/2}}$ is the number of basis states at half-filling). Figure \ref{fig:timeEvo} shows the time dependence of $D$, at fixed system size $N=12$ and $\Omega/J=0.83$, for large and small values of interaction (the same as those in Figure \ref{fig:spectrum}).  The $C$ state, which begins with a thermal $D$ value, stays as such during time evolution. However, the evolution of the $A$ and $B$ states, which begin with non-thermal values of $D$, remain non-thermal at large interaction strength with quite small temporal fluctuations. This memory of the initial doublon density at long times suggests that the Floquet eigenstates have significant overlap with states of definite doublon number, although one cannot definitively conclude this on the basis of finite time data alone as the possibility of prethermalization exists. This supports our intuition that doublons are indeed an appropriate characterization of physics in this model and are a useful signature of non-thermality. In contrast, for small interaction, resonances efficiently mix doublon-like states and all initial conditions evolve to a thermal $D$ value.

\section{\label{sec:Scaling} Scaling}

\begin{figure}
\includegraphics[width = 3.5in]{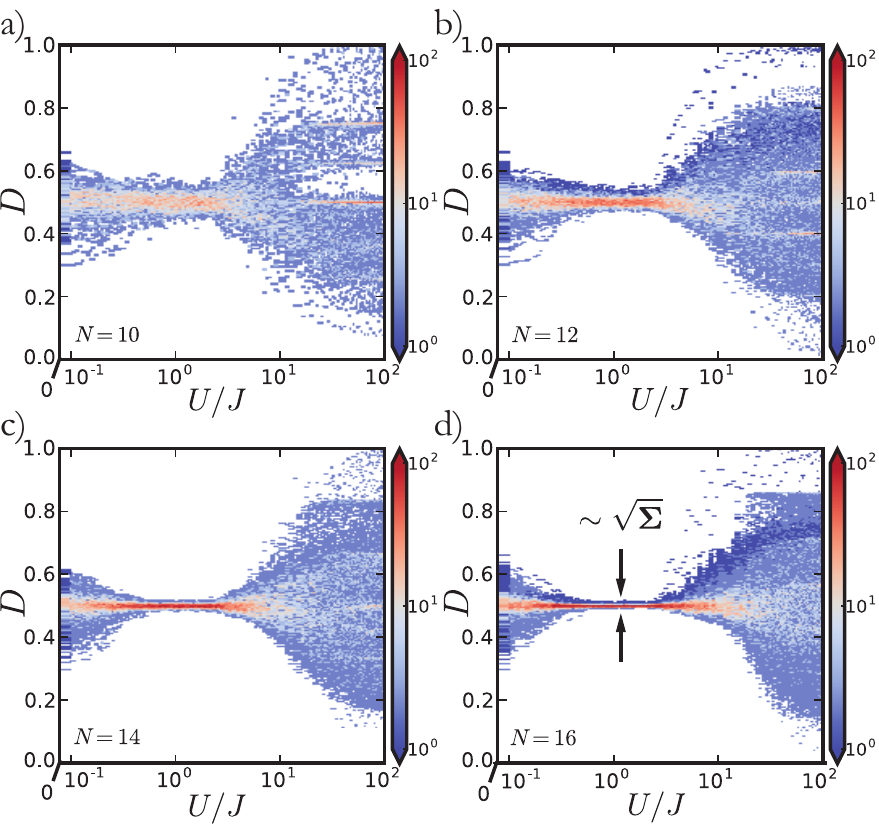}
\caption{Histogram of $D_n$ measured in each Floquet eigenstate as a function of $U/J$ and system size. For $U/J\ll1$, the spectrum displays some spread in the doublon density due to  near-integrability close to the free fermion limit $U=0$. At $U/J\sim O(1)$, however, sufficient mixing leads to a tight squeeze of $D$ around $0.5$, indicating a thermal region. At strong interactions $U/J \gg 1$, there is significant spread of $D$ also indicating non-thermal behavior.}
\label{fig:histogram}
\end{figure}

In this section, we explore the dependence of the doublon density on system size and interaction strength at a fixed frequency $\Omega/J=0.83$. We show that two different regimes, characterized as non-thermal and thermal, arise, each with distinct scaling behavior of the spectral variance of the doublon density. The two regimes are separated by a crossover in interaction strength that has power-law dependence in system size. 

To understand these statements, let us first consider the histogram of $D$ over all Floquet eigenstates in the spectrum as a function of the coupling $U/J$ (Figure \ref{fig:histogram}). Interestingly, at large values of interaction, $U/J\gg1$, the Floquet spectrum exhibits a large variance in the values of $D$ characteristic of non-thermal behavior due to integrable behavior in the $U/J\rightarrow\infty$ limit (see Section \ref{sec:Integrability and its breakdown} for detailed discussion). As the interaction decreases to $U/J\sim O(1)$ , there is a tight clustering of $D$ values around $0.5$. This infinite temperature thermal behavior is due to the heating and mixing expected from that of a generic closed driven interacting system. For very small values of interaction, $U/J\ll1$, the spectrum has some doublon variance close to that of a purely static metallic spectrum ($U\rightarrow0$). This is precisely the same type of finite-size non-thermal behavior manifesting itself around the free fermion integrable point. Near this point, however, doublons are not the ideal observable suited to gauging non-thermality and so the deviations away from the infinite temperature value are weak. We will term the situation when non-thermality arises due to finite size as near-integrability. 

As the system size $N$ increases, we see that the thermal region gets more tightly centered around the infinite temperature value and persists to stronger interaction. Moreover, the near-integrability region governed by free fermions shrinks closer to $U/J = 0$. Therefore, extrapolating to the thermodynamic limit, we conclude that the entire system is likely in a thermal phase for any nonzero finite interaction strength. This is precisely the usual infinite temperature scenario for a generically non-integrable Floquet system. Regardless of the featureless thermodynamic limit, however, Figure \ref{fig:histogram} demonstrates that small system sizes host non-thermal regimes.

\begin{figure}
\includegraphics[width = 3.5in]{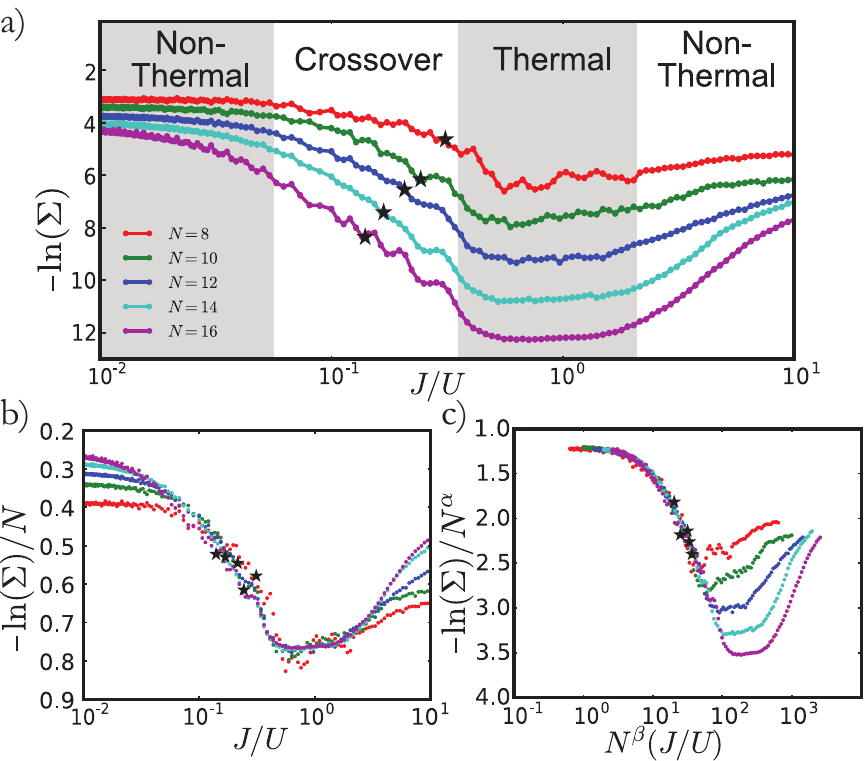}
\caption{Dependence of doublon log spectral variance on coupling and and system size. Figure a) shows raw data which demonstrate the three regions clearly, near-integrability for $J/U\gg1$, thermal for $J/U \sim O(1)$, and non-thermal (also near-integrability) for $J/U\ll1$. The black stars indicate the approximate midpoint of the crossover region. Figure b) rescales the axes to show the scaling collapse of the thermal region indicating simple exponential behavior independent of coupling. Figure c) rescales the axes differently to show the scaling collapse of the non-thermal region with $\alpha=0.45$ and $\beta=2.0$.}
\label{fig:scaling}
\end{figure}

The spread of the distribution of the doublon density characterizes the non-thermality of the system at a particular interaction strength. In line with this expectation, we calculate the log spectral variance $\mathrm{ln}(\Sigma)$ of $D$ for various system sizes as a function of $J/U$ in Figure \ref{fig:scaling}a. The variance clearly indicates each of the regions discussed above: the free fermion near-integrability region for $J/U\gg1$, the thermal region for $J/U\sim O(1)$ with the smallest variances, the crossover region with midpoints denoted by black stars, and the non-thermal region $J/U\ll1$ with the largest variances (also a near-integrability region). Note that black stars representing the crossover region are not uniquely defined. Here, we choose them to be close to the midpoint between the average log spectral variance values in the non-thermal and thermal regions.

We can distinguish the thermal and non-thermal regimes quantitatively by observing their distinct scaling forms (see Eq. \ref{eq:scalingForms}). In Figure \ref{fig:scaling}b, we see that the variance has simple exponential decay in system size with no dependence on interaction. In contrast, Figure \ref{fig:scaling}c shows that the non-thermal regime has a non-trivial scaling function (denoted by $f$ in Eq. \ref{eq:scalingForms}) with joint dependence on system size and interaction.

\begin{eqnarray}
\Sigma & \sim & \begin{cases}
e^{\kappa N},(\kappa=-0.77) & \mathrm{Thermal}\\
e^{N^{\alpha}f(N^{\beta}\frac{J}{U})},(\alpha=0.45,\beta=2.0) & \mathrm{Non-thermal}
\end{cases}
\label{eq:scalingForms}
\end{eqnarray}

Taking the midpoints of the crossover region as an approximate ``phase boundary," we obtain the power-law

\begin{eqnarray}
\p{\frac{J}{U}}_{c} & \approx & 2.9N^{-1.1}
\label{eq:crossoverScaling}
\end{eqnarray}
shown in Figure \ref{fig:phasediagram}. The power-law exponent for the crossover may be understood as the intermediary behavior between the limits given by the two scaling forms - the non-thermal region suggests a crossover dependence of $N^{-\beta}$ while the thermal region suggests no system size dependence. As expected, the non-thermal region seems to vanish in the thermodynamic limit at fixed values of the couplings and drive, but there is still a non-trivial dependence on system size that suggests that heating will not take place given the appropriate order of limits.

\section{\label{sec:Integrability and its breakdown} Integrability and its breakdown}

The source of the non-thermal regime at large interaction strength is the integrability of the system in the limit $U/J\rightarrow\infty$. Note that this integrable limit is \textit{not} the same the $J=0$ integrable point, a distinction that will become clear shortly. In this section, we discuss the $U/J\rightarrow\infty$ integrable limit of the Floquet Hamiltonian, following which we analyze the breakdown of integrability in a perturbative expansion in $J/U$. Finally, we also discuss the onset of the infinite temperature thermal phase from the perspective of this expansion.

Effective Hamiltonians for Floquet systems are often obtained by perturbative methods, treating inverse frequency, $\Omega^{-1}$, as a small parameter. In our system, we explicitly consider the highly resonant regime at strong drive strength, so direct application of high frequency expansions (HFE) such as Magnus or van Vleck is invalid. Rather, to obtain a controlled expansion in the limit of large interaction strength, it is convenient to go into a frame rotating with the driven interaction term, similar to that used for the Fermi-Hubbard model in Ref.~\onlinecite{Bukov_PRL_2016,MentinkEckstein2015,GorgEsslinger2017,KitamuraAoki2017,NieOshikawa2013}. In this rotating frame, the Fourier harmonics of the Hamiltonian come with sharply peaked coefficients, which upon use in the van Vleck expansion, yields a controlled expansion in $J/U$. Let us see how this procedure works. 

Consider changing frame via the unitary transformation, $V(t) = \exp\left(-i\kappa F(\Omega t)\sum_{j}n_{j}n_{j+1}\right)$, where $F(\Omega t)=\int f_U(t) d(\Omega t)$. This drive is chosen to cancel the bare interaction term and replace it by strong oscillations of the dressed hopping term. This gives the rotated Hamiltonian $\tilde H = i\d_{t}(V^{\dag})V+V^{\dag}HV$, with

\begin{eqnarray}
\tilde{H} & = & \sum_{m=0,\pm1}\tilde{H}_{m}e^{im\kappa F(\Omega t)}\\
\tilde{H}_{0} & = & J\sum_{j}\delta_{n_{j-1},n_{j+2}}(c_{j}^{\dag}c_{j+1}+c_{j+1}^{\dag}c_{j}) \label{eq:Htilde_0}\\
\tilde{H}_{1} & = & J\sum_{j}\p{n_{j-1}(1-n_{j+2})c_{j}^{\dag}c_{j+1}+n_{j+2}(1-n_{j-1})c_{j+1}^{\dag}c_{j}}\nonumber\\
\tilde{H}_{-1}& = &\tilde{H}_{1}^{\dag}
\end{eqnarray}
where $\kappa=U/\Omega$ and $\delta_{n_{j-1},n_{j+2}} = (1-n_{j-1}-n_{j+2}+2n_{j-1}n_{j+2})$ is a constraint which allows nearest neighbor hopping only if both adjacent sites are either occupied or unoccupied. Decomposing the rotated Hamiltonian into harmonics, $\tilde{H} = \sum_{l}e^{il\Omega t}\tilde{H}^{(l)}$, we obtain the relation $\tilde{H}^{(l)} = \sum_{m=0,\pm1}\tilde{H}_{m}\alpha_{l}(m\kappa)$ where $\alpha_l$ are the Fourier coefficients of the rotating frame drive: $e^{im\kappa F(\Omega t)} = \sum_{l}e^{il\Omega t}\alpha_{l}(m\kappa)$. Importantly, for a square wave drive, $\alpha_l(m\kappa)$ is peaked to a constant of order 1 around $l=\pm m\kappa$ and quickly decays away from this point, a crucial property for our approach which will exist much more generally than just the square wave considered here. Performing the HFE in this frame produces an effective Hamiltonian $H_{\mathrm{eff}} = H_{\mathrm{eff}}^{[0]}+H_{\mathrm{eff}}^{[1]}+H_{\mathrm{eff}}^{[2]}+...$ with terms $H_{\rm eff}^{[n]} \sim \Omega^{-n}$ that do not seem to appear small. The fact that $\alpha_l(m\kappa)$ is sharply peaked counteracts the inverse frequency coefficient precisely in a way so as to yield an approximate $J/U$ expansion. Therefore, even though we are not in the limit of large frequency, the expansion is physically meaningful. By performing an appropriate rotation of the effective Hamiltonian computed up to $n$-th order, we obtain an approximate stroboscopic Floquet Hamiltonian $H_F^{[n]}$. More details on the rotating frame and subsequent high frequency expansion may be found in Appendix \ref{sec:Expansion}.

The leading order term of the HFE in the limit $\kappa\rightarrow \infty$ yields $H_F^{[0]}=H_{\rm eff}^{[0]}=\tilde{H}_0$. This corresponds to the time-independent correlated hopping model arising from the aforementioned constraint. Note that this is quite interesting since the $U\rightarrow \infty$ limit yields a non-trivial correlated hopping model, quite different from the case of $J=0$ which, in the rotated frame, would yield $\tilde H(J=0) = 0$. Furthermore, the $J=0$ Hamiltonian has locally conserved doublon numbers while the correlated hopping model in $U\rightarrow\infty$ only has a globally conserved doublon number, though as we will see shortly, it is still an integrable model. Higher order corrections such as $H_\mathrm{eff}^{[2]}$ break both this global doublon number symmetry and integrability as discussed briefly below and in more detail in Appendices \ref{sec:Expansion} and \ref{sec:Evidence}.

Let us now discuss the integrability of the correlated hopping Hamiltonian, $\tilde{H}_0$ defined in Eq. \ref{eq:Htilde_0}. A priori it is not obvious that $\tilde{H}_0$ maps into an integrable Hamiltonian\textcolor{blue}. The Hilbert space of $\tilde{H}_0$ are states with fermions at half-filling. Let us start by mapping the Hilbert space to states defined on its dual-lattice, given by the position of the domain walls which separate an occupied region from an unoccupied one. For example, on $10$ sites \footnote{We assume open boundary conditions throughout the whole paper. In this case, there is no distinction between fermions and hardcore bosons. However, if considering the system on a ring, then one must be careful about (anti-)periodic boundary conditions as exchange statistics are relevant \cite{NieOshikawa2013}.},
 \begin{equation}
 |0011111000\rangle \rightarrow |0d0000d00\rangle
 \end{equation}
It is possible to rewrite the constrained hopping processes as nearest neighbor hopping of pairs of domain walls,
\begin{eqnarray}
|\cdots 1011 \cdots \rangle \leftrightarrow |\cdots 1101\cdots \rangle \equiv |\cdots dd0\cdots\rangle \leftrightarrow |\cdots 0dd\cdots \rangle, \nonumber\\
|\cdots 0010 \cdots \rangle \leftrightarrow |\cdots 0100\cdots \rangle \equiv |\cdots 0dd\cdots\rangle \leftrightarrow |\cdots dd0 \cdots \rangle, \nonumber
\end{eqnarray}
with the constraint that the domain walls, $d$, are hardcore particles. Note that flipping $1\leftrightarrow 0$ in the original fermions, maps to the same state of domain walls. This is a result of a particle-hole symmetry of $\tilde{H}_0$ in the language of the bare fermions. Also note that the correlated hopping conserves the total number of doublons. Therefore, the doublon spectral variance $\Sigma$, and indeed the full counting statistics of the doublon number, may be readily obtained in the $U\to \infty$ limit\footnote{This is acheivable using the probability of finding a $k$-doublon state of a half-filled $N$ site system given by $p_N(k)={{\frac{N}{2}+1}\choose{k+1}}{{\frac{N}{2}-1}\choose{k}}$ as in Eqn. \ref{eq:DoublonInf}. }.

This pair hopping of domain walls can be further mapped to free fermions. To do so, we map the basis states of the domain walls, denoted as a string of $d$s and $0$s, into those of a new particle $\tilde{d}$ in a truncated Hilbert space as follows:
\begin{enumerate}
\item If a site is unoccupied, leave it alone: $0\rightarrow 0$
\item  Given a string of $d$'s, replace them pairwise by $\tilde d$'s: $d\to\emptyset$, $dd\to \tilde{d}$, $ddd\to \tilde{d}$, $dddd\to \tilde{d}\tilde{d}$ and so on.
\end{enumerate}
This second step comes from noting that an isolated $d$ particle is essentially frozen, such that a pair of $d$s can hop right through it, or equivalently the $d$ particle reassociates into a new pair. Thus isolated $d$s play no dynamical role, and may be removed from the Hilbert space. We note here that a similar mapping to free fermions from repulsive nearest-neighbor interacting fermions has been done in Ref. \onlinecite{Henley_2009}. However, it remains an open question as to whether more general constrained hopping models are integrable.

Interestingly, the above mapping takes several different states of $d$s to the same state of $\tilde{d}$s. This is a hidden symmetry in $\tilde{H}_0$ and gives rise to massive degeneracy in its energy spectrum. Since the $\tilde{d}$ particles behave like a pair of domain walls, the Hamiltonian $\tilde{H}_0$ in this new basis is just free particle hopping with matrix element $J$, i.e., $\tilde{H}_{0} = J\sum_{i}\tilde{d}_{i}^{\dag}\tilde{d}_{i+1}+h.c.$. This is the origin of the integrability when $U \rightarrow \infty$ keeping $J$ finite. We expect that the long-time limit of the non-thermal regime is smoothly connected to this $U\rightarrow\infty$ free fermion integrable point. Therefore we hypothesize that the long-time state in the non-thermal regime is well-described by a time-periodic GGE \cite{Lazarides_Moessner_2014_PRL,RussomannoSantoro2012,Vidmar_Rigol_GGE_Review,PolkovnikovColloquium11} from the perspective of local observables. An explicit check is the subject of future work.

Having understood integrability of the infinite $U$ case, we can now briefly discuss its breakdown at finite $U$ and how this behavior changes as a function of system size. At finite large $U$, there are additional contributions to $H_{\rm eff}$. For example, even at zeroth order in the HFE (see Eq. \ref{eq:VVE-full}), there are additional contributions from $\alpha_0(\pm \kappa) \tilde{H}_{\pm 1}$. As discussed in Appendix \ref{sec:Expansion}, the terms in $\tilde{H}_{\pm 1}$ result in pair-creation/annihilation of $d$ particles. At second order (the first order term vanishes by symmetry of the drive), higher  harmonics contribute to the effective Hamiltonian. As a result, this mapping to free $\tilde{d}$ particles breaks down. Thus, higher order terms may break integrability while keeping the HFE convergent. However, an alternative mechanism also exists. For a given finite $U$, the HFE itself may be invalid (or inaccurate), possibly at all orders, which certainly would break the infinite $U$ integrability. 

In principle, the breakdown of integrability due to higher order terms and the breakdown of the HFE can occur with distinct system size dependence. One can envision two possible scenarios for the crossover from integrable dynamics at $U\rightarrow\infty$ to an infinite temperature Floquet-ETH phase at finite $U$,

\begin{enumerate}

\item Integrable, \fbox{Non-Thermal} $\rightarrow$ Non-Integrable, \fbox{Floquet-ETH} 

\item Integrable, \fbox{Non-Thermal} $\rightarrow$ Non-Integrable, \fbox{Finite Temperature ETH} $\rightarrow$ Non-Integrable, \fbox{Floquet-ETH}

 \end{enumerate}

\begin{figure*}

\includegraphics[width = \textwidth]{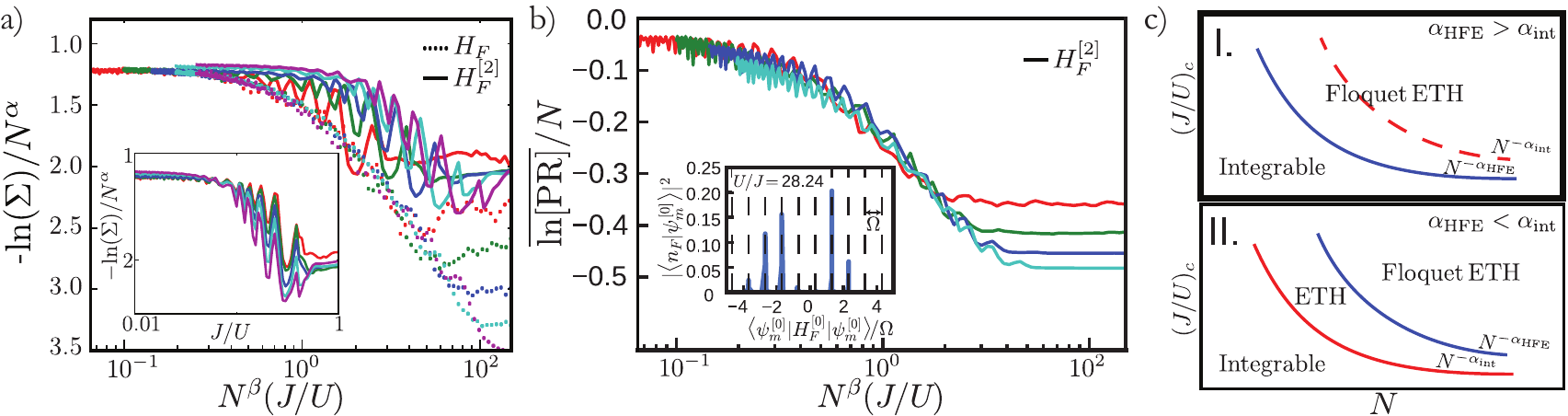}
\caption{Dependence of the doublon log spectral variance on coupling and system size for both $H_F$ and $H_F^{[2]}$.  Figure a) compares the data from Figure \ref{fig:scaling}c to the same data gathered from $H_F^{[2]}$. Note that $H_F^{[2]}$ has a different scaling form as shown in the inset, displaying a much weaker dependence on system size. Breakdown of the HFE happens faster than the breakdown of integrability within the HFE, apparently resulting in a direct transition from integrability to an infinite temperature Floquet-ETH phase. Figure c) depicts this scenario in the bold top box while displaying an alternative possibility in the bottom box which contains an intermediate finite-temperature ETH phase. Figure b) displays the average log participation ratio (LPR) of exact Floquet eigenstates in the basis of zeroth-order HFE eigenstates. Note that the LPR has the same scaling form as the log spectral variance and is a good measure of delocalization (here due to resonances) of exact Floquet eigenstates in the basis of zeroth-order HFE states. An explicit example of this is shown in the inset for a representative exact Floquet eigenstate.}
\label{fig:FloquetvsVVE}

\end{figure*}

The second scenario is plausible when a non-integrable effective Hamiltonian is obtained from a convergent HFE. With these two cases in mind, we examine the variance of the doublon density, comparing that of $H_F^{[2]}$ with that of the exact Floquet Hamiltonian. As shown in Figure \ref{fig:FloquetvsVVE}a, the variance 
data from $H_F$ has a scaling collapse for the non-thermal plateau that  breaks down at an earlier point than that of the non-thermal region predicted by $H_F^{[2]}$. In fact, as shown in the inset, $H_F^{[2]}$ exhibits a different scaling form for the non-thermal plateau. Therefore, our data indicate that the breakdown of the HFE -- yielding a Floquet-ETH phase -- happens first, corresponding to the first scenario. Hence, we do not observe any physics corresponding to a finite temperature ETH states. If we assume the finite-size breakdown of integrability within the HFE and of the HFE itself correspond to distinct finite-size scaling laws $(J/U)_c \sim N^{-\alpha_\mathrm{int}}$ and $(J/U)_c \sim N^{-\alpha_\mathrm{HFE}}$ respectively, as depicted in Figure \ref{fig:FloquetvsVVE}c, then our data indicates that $\alpha_\mathrm{HFE}>\alpha_\mathrm{int}$. If instead we had $\alpha_\mathrm{HFE}<\alpha_\mathrm{int}$, we would be able to achieve the second scenario where a finite temperature ETH regime emerges between the integrable and infinite temperature Floquet-ETH phases.

Finally, the breakdown of the HFE provides an explanation for the physical mechanism of thermalization. There remain two possible routes to the breakdown of the HFE. The first is the breakdown of the operator expansion, whereby the magnitude of the higher order terms relative to the zeroth order term becomes significant. In Appendix \ref{sec:Evidence}, we examine the behavior of the trace-norm of the second order term, $H_{\rm eff}^{[2]}$ in comparison with the zeroth order term, $H^{[0]}_{\rm eff}$. As shown in Figure \ref{fig:tracenorm} (Appendix \ref{sec:Evidence}), this does not capture the finite size scaling of the crossover region. While for small $J/U$ the trace-norm of $H_\mathrm{eff}^{[2]}$ increases as a function $(J/U)^2$, it has no finite-size dependence which is inconsistent with the scaling observed in the distribution of the doublon density. This rules out a breakdown of the operator expansion, at least at second order. 

The second route is through a proliferation of resonances \cite{Bukov_Polkovnikov_2016_PRB}. This proliferation is analogous to a localization-delocalization transition in the space of many-body eigenstates of $H_{\rm eff}^{[0]}$ (denoted by $|\psi_{m}^{[0]}\rangle$). When $J/U\rightarrow 0$, the exact Floquet eigenstates (denoted by $|n_F\rangle$) are identical to $|\psi_{m}^{[0]}\rangle$, corresponding to a localized state. As $J/U$ increases, the drive induces resonances with states energetically separated by $\Omega$ such that the eigenstates of $H_{\rm eff}^{[0]}$ cease to faithfully represent the Floquet eigenstates due to non-perturbative instanton-like effects. It has been argued \cite{Bukov_Polkovnikov_2016_PRB,Weinberg_Kolodrubetz_2017} that in fact no finite-order HFE eigenstates capture these resonances, which is consistent with our results for $H_F^{[2]}$ (not shown). Therefore, when these resonances become active, the HFE completely breaks down. We can quantify the breakdown by viewing the  proliferation of resonances as a delocalization of the exact Floquet states in the space of the zeroth-order HFE eigenstates,$\{|\psi_{m}^{[0]}\rangle\}$. This property is nicely characterized by the spectrum-averaged log participation ratio (LPR), $\overline{\ln[\mathrm{PR}]} = \mathrm{Dim}[\s H]^{-1}\sum_{n_{F}}\ln\p{\sum_{m}|\qb{n_{F}}{\psi_{m}^{[0]}}|^{4}}$, shown in Figure \ref{fig:FloquetvsVVE}b. With increasing $J/U$, the participation ratio decreases, indicating eigenstate delocalization. The scale at which the LPR plateaus roughly agrees with the scale at which the eigenstates appear to be thermal. Furthermore, the system size scaling is consistent with that of the doublon density. This strongly indicates that in our system, the proliferation of resonances is responsible for the breakdown of the HFE. The inset in Figure \ref{fig:FloquetvsVVE}b shows an explicit example of the appearance of such resonances, which are already active at a relatively strong drive $U/J=28.24$.

In summary, we have shown that the non-thermal behavior of the driven Hamiltonian at large $U/J$ can be traced back to the integrability of the $U\rightarrow \infty$ point, where the HFE gives the effective description of the Floquet eigenstates. At finite system sizes, non-thermal behavior is observed at a large but finite $U/J$. The crossover from the integrable-to-thermal behavior of the eigenstates as a function of $U/J$ is governed by the proliferation of resonances induced by the drive. The finite size scaling of such resonant breakdown is numerically consistent with the finite size scaling of the doublon density, a fact which remains to be understood analytically.

\section{\label{sec:Discussion} Discussion and Conclusions}

We have studied a strongly-driven system of interacting spinless fermions and found an unexpected non-thermal regime at large interaction strength and finite system size.
We have shown that this non-thermal regime is due to the integrability of the system at infinite $U$ that weakly persists to large but finite $U$ at finite size, a phenomenon that we call near-integrability. We found power-law scaling of the crossover region, i.e. where the system goes from integrable to non-integrable, with system size. We argued that this crossover comes from a breakdown of the high-frequency expansion leading immediately to an infinite temperature Floquet-ETH phase with no intervening finite temperature regime for our choice of parameters. Further evidence for the qualitative independence of these phenomena upon the details of the model may be found in the appendices.

Our analysis from the effective $J/U$ expansion indicates the intriguing possibility of a periodically driven system in which integrability is first broken to a finite temperature ETH phase before breaking down to the infinite temperature Floquet-ETH phase. This scenario seems plausible and is quite interesting in that it runs counter to the commonly held intuition that isolated, periodically driven interacting systems heat to infinite temperature. Reference \onlinecite{ClaeysCaux2017} studies integrability breakdown in a driven Heisenberg chain as one moves away from high frequency limit. In a certain parameter regime, they find evidence for such a finite temperature ETH (as well as another regime where resonant breakdown occurs). In the present model, such a phase is expected to arise when $J \ll \Omega \ll U$ (or perhaps less interestingly, at even higher frequencies $\Omega \gg U \gg J$) such that resonances vanish while the interactions still strongly influence the states. Future work to explore such a intermediate phase and connect it to related finite time phenomena such as prethermalization \cite{ElseNayak2017PRX,AbaninHuveneers2017,Weidinger_Knap_2017} remains an ongoing challenge. 

Our results are immediately relevant to a wide variety of engineered quantum systems, 
where finite system size is currently a given. Even in larger systems, our finite size scaling should provide insight into the local thermalization dynamics of finite size subsystems, which may be coarse grained towards understanding the larger-scale thermalization dynamics of the full system. This is deeply related to time scales for prethermalization, in which the dynamics is dominated by the nearby integrable point \cite{BergesWetterich2004,KolathAltman2007,EcksteinPhillip2009,MoeckelKehrein2010,MatheyPolkovnikov2010,BarnettVengalattore2011,GringSchmiedmayer2012}. 

We can estimate the scaling of the prethermalization time $t^\ast$ for an infinite system by assuming that a finite subsystem appears thermal with respect to local observables when $N \geq N^\ast$ due to sufficient mixing. Noting that $J$ sets the characteristic velocity in the model, we may approximate the prethermalization time as $t^\ast \sim N^\ast/J = (U/J^{3})^{1/2}$ using the scaling behavior at the edge of the non-thermal region.

We expect that understanding the finite size and finite time scaling in a more rigorous way -- as done in this work for one model -- will allow better understanding of heating mechanisms. This in turn should allow control of heating, which is a crucial step for the experimental realization of novel Floquet phases that are able skirt their boring infinite temperature fate. While preparing this manuscript, the authors became aware of upcoming complementary work by by Peronaci et al.\cite{Schiro}.

\textit{Acknowledgements}. The authors would like to thank Jim Garrison, Marin Bukov, Anatoli Polkovnikov, Evert van Nieuwenburg, Yuval Baum, Min-Feng Tu and Joel Moore for insightful discussions. MK thanks funding from the Laboratory Directed Research and Development from Berkeley Laboratory, provided by the Director, Office of Science, of the U.S. Department of Energy under Contract No. DEAC02-05CH11231, and from the U.S. DOE, Office of Science, Basic Energy Sciences as part of the TIMES initiative. GR and KS are grateful for support from the NSF through DMR-1410435, the Institute of Quantum Information and Matter, an NSF Frontier center funded by the Gordon and Betty Moore Foundation, the Packard Foundation, and from the ARO MURI W911NF-16-1-0361 ``Quantum Materials by Design with Electromagnetic Excitation" sponsored by the U.S. Army. KS is additionally grateful for support from NSF Graduate Research Fellowship Program. P.T. is supported by National Research Council postdoctoral fellowship, and acknowledges funding from ARL CDQI, NSF PFC at JQI, ARO, AFOSR, ARO MURI, and NSF QIS.

\appendix

\section{Frequency Dependence \label{sec:Frequency Dependence} }

\begin{figure}
\includegraphics[width = 3.5in]{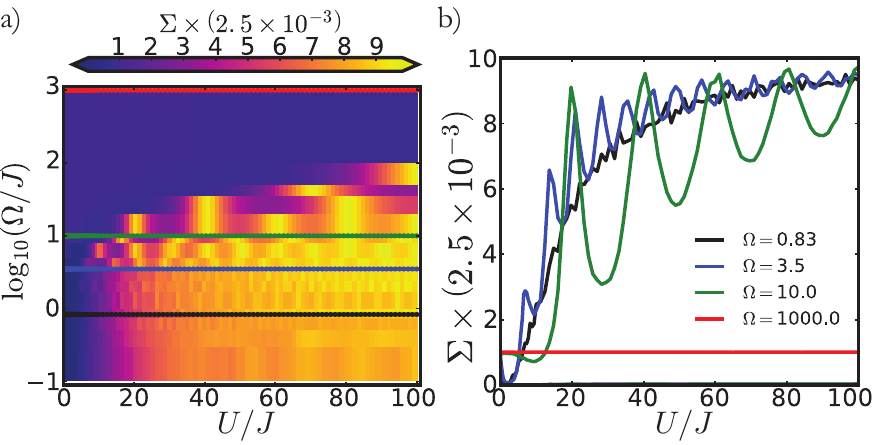}
\caption{Frequency dependence of spectral doublon variance as a function of coupling $U/J$ at $N=12$. Figure a) shows frequency along the y-axis and coupling along the x-axis with color denoting the variance value. Figure b) shows cuts at particular frequencies. In the high frequency limit, the system is approximated by the time-averaged lab frame Hamiltonian, leading to a variance given by static free fermions. In the low-frequency limit, we get the variance behavior discussed in the text which shows the thermal to non-thermal transition as $U/J$ gets larger. At intermediate frequencies, the rare resonances govern the precise details of the variance (e.g.\mkadd{,} peaking) and the system is quite sensitive to drive parameters.}
\label{fig:frequency}
\end{figure}

In this appendix, we consider a fixed system size $N=12$ and study how the spectral variance changes as a function of frequency and interaction strength. Previously, we discussed the variance properties for the highly resonant case at low frequencies. In the opposite limit, at very large frequencies surpassing the many-body bandwidth, the system can be effectively described by the time-averaged Hamiltonian. For our model, the time-averaged Hamiltonian is just free fermions with nearest-neighbor hopping. Therefore, at very large frequencies, we expect the variance to be the same as that of a static metal with no dependence on interaction. Indeed, we see these two limits in Figure \ref{fig:frequency}.

For intermediate frequencies, where state mixing due to resonances is weaker, the variance shows peaks at even integer values of $U/\Omega$. This is due to the fact that the square wave contains only odd harmonics of $\Omega$. At odd multiples of $\Omega$, the system has an additional resonance contributing to mixing thereby decreasing the variance closer to its thermal value. The peaks at even integer values of $U/J$ are precisely the opposite situation where these extra processes are most energetically suppressed consequently resulting in weaker mixing. We have checked that indeed choosing different waveform compositions changes this peaking phenomena accordingly (not shown). The conceptual point here is that in the intermediate frequency regime, the system is quite sensitive to the rare resonances that occur and hence the precise details of the spectrum and drive carry serious impact on its the thermalization properties. Overall, however, even if resonances are weaker, the same general onset of non-thermal behavior with increasing interaction exists.

\section{\label{sec:Waveform Dependence} Waveform Dependence}

\begin{figure}
\includegraphics[width = 3.5in]{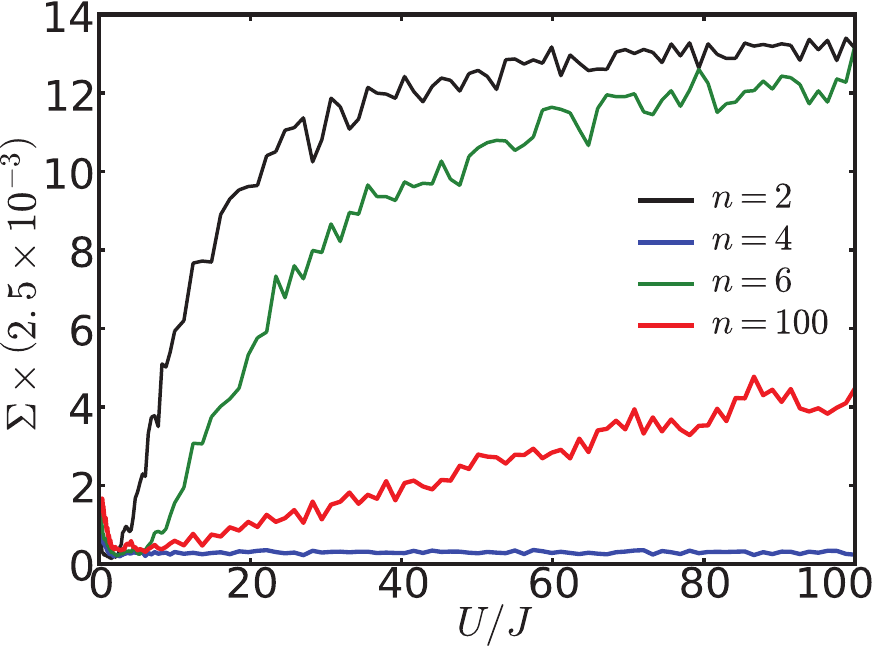}
\caption{Waveform dependence of spectral doublon variance as a function of coupling $U/J$ at $\Omega/J=0.83$. A square wave is given by $n=2$ and a cosine is closely approximated by $n=100$. In between, the discretization of sampling a waveform gives rise to dampening and resurgence effects as can be understood by considering the time-evolution operator over one period $U(T,0)$ (see text). Overall, the thermal to non-thermal transition persists for a cosine drive but has significantly slower crossover behavior as compared to the square drive.}
\label{fig:waveform}
\end{figure}

In this section, we work in the highly resonant regime at fixed system size $N=10$ and discuss how the choice of waveform can alter the behavior of the thermal to non-thermal transition. We can study this systematically by introducing a parameter $n$ that denotes the number of steps the waveform takes in approximating a single cosine harmonic over a period, i.e., we discretize the cosine function in time with $n$ steps and the amplitude of the $j$-th step given by $\mathrm{cos}(\frac{2\pi j}{n})$ for $j=0,1,\ldots,n-1$. The case of $n=2$ corresponds to the square wave. As $n\rightarrow\infty$, we obtain a perfect cosine function. In between, we may track how continuous interpolation between a square wave and a single harmonic affects the variance. 

Figure \ref{fig:waveform} shows four cases of how the variance changes with increasing $n$. Upon increasing $n$ from $2$ to $4$, we see a sudden drop of the variance. Again increasing $n$ from $4$ to $6$ results in a resurgence of the variance. Finally, at $n=100$ where we well-approximate a cosine drive, the variance grows roughly linearly as a function of interaction. 

The intuition for this seemingly odd behavior is apparent by considering the time evolution operator over one period $U(T,0)$. The unitary $U(T,0)$ contains Hamiltonians $H_j$ constructed from the discrete cosine amplitudes. For $n=2$, the interaction contributes terms with amplitudes $U,-U$ for time steps of $T/2$ and so $U(T,0)$ spends all its time with the interaction at $|U/J|$. In contrast, for $n=4$, the interaction contributes terms with $U,0,-U,0$ for time steps of $T/4$. In this case, we see that for half the time period, $U(T,0)$ contains evolution due to a purely static non-interacting metal. Intuitively speaking, this severely weakens the ``effective" interaction scale over a period and thus leads to a more thermal variance than the case of $n=2$ where the $U=0$ values are absent. Upon further increasing the sampling to $n=6$, the interaction steps no longer contain the $U=0$ value and hence the variance returns to a larger value. Of course, however, at fixed $U/J$, $n=6$ indeed has a weaker effective interaction scale than that of a square wave and so the variance, while still demonstrating the same overall trend to non-thermality with increasing interaction, is dampened. This trend saturates apparently with approximately linear growth of variance with $U/J$ for a single harmonic at $n=100$. All of this suggests that even though a single harmonic contains contributes fewer resonances than a square wave, which apriori one might expect to lead to more non-thermal behavior, the fact that the effective interaction scale is greatly reduced at fixed $U/J$ for a single harmonic dominates the thermal to non-thermal crossover behavior.

\section{Derivation of the Effective $J/U$ Expansion \label{sec:Expansion} }
In this section, we provide the derivation of an effective $J/U$ expansion derived from a van Vleck high frequency expansion (HFE), though we are explicitly not working at high frequency. We move to a rotating frame which eliminates the interaction term via the unitary transformation
\begin{eqnarray}
\tilde{H} & = & i\d_{t}(V^{\dag})V+V^{\dag}HV\nonumber\\
\kett{\tilde{\psi}} & = & V^{\dag}\kett{\psi}\nonumber\\
V(t) & = & e^{-i\kappa F(\Omega t)\sum_{j}n_{j}n_{j+1}}\mkadd{,}
\label{eq:rotation}
\end{eqnarray}
where $\kappa=U/\Omega$, and $F(\Omega t)$ is the integral of the drive with respect the variable $\Omega t$. This yields the transformation of the annhilation operator $\tilde{c}_{i} = V^{\dag}(t)c_{i}V(t) = e^{-i\kappa F(\Omega t)(n_{i-1}+n_{i+1})}c_{i}$ which can be immediately used to construct the rotated Hamiltonian
\begin{eqnarray}
\tilde{H} & = & J\sum_{i}(\tilde{c}_{i}^{\dag}\tilde{c}_{i+1}+h.c.)\nonumber\\
 & = & J\sum_{i}(e^{i\kappa F(\Omega t)(n_{i-1}-n_{i+2})}c_{i}^{\dag}c_{i+1}+h.c.)
\label{eq:Hrot}
\end{eqnarray}
Note that the time-dependence of the rotated Hamiltonian disappears if $n_{i-1}=n_{i+2}$, a property which will lead to interesting results. The above form suggests a convenient expansion $\tilde{H} = \sum_{m=0,\pm1}\tilde{H}_{m}e^{im\kappa F(\Omega t)}$ upon factoring out the operator content in the exponential in \eqref{eq:Hrot}.
\begin{eqnarray}
\tilde{H}_{0} & = & J\sum_{j}\delta_{n_{j-1},n_{j+2}}(c_{j}^{\dag}c_{j+1}+c_{j+1}^{\dag}c_{j})\nonumber\\
\tilde{H}_{1} & = & J\sum_{j}\p{n_{j-1}(1-n_{j+2})c_{j}^{\dag}c_{j+1}+n_{j+2}(1-n_{j-1})c_{j+1}^{\dag}c_{j}}\nonumber\\
\tilde{H}_{-1} & = & \tilde{H}_1^{\dag},
\end{eqnarray}
where $\delta_{n_{j-1},n_{j+2}} = (1-n_{j-1}-n_{j+2}+2n_{j-1}n_{j+2})$ is a constraint which allows nearest-neighbor hopping only if $n_{i-1}=n_{i+2}$, i.e.\mkadd{,} the adjacent sites have the same density. This type of correlated hopping preserves total doublon number. In sharp contrast, $\tilde{H}_{\pm 1}$ allows nearest-neighbor hopping only if $n_{i-1}\ne n_{i+2}$ and therefore can be understood as doublon creation and annihilation. Hence, these terms explicitly break the global doublon number symmetry. If one were to think about this correlated hopping in terms of domain wall dynamics on the bonds of the lattice, $\tilde{H}_{0}$ would be responsible for nearest-neighbor hopping of domain wall pairs (see section \ref{sec:Integrability and its breakdown} and appendix \ref{sec:Evidence}) while $\tilde{H}_{\pm 1}$ would be responsible for domain wall pair creation and annihilation. This intuitive understanding suggests that in the limit of $U \rightarrow \infty$, where only $\tilde{H}_{0}$ is active on average, $\tilde{H}$ is integrable. Formalizing this intuition mathematically is rather tough, but we discuss an algorithm for checking integrability in appendix \ref{sec:Evidence}.

\begin{figure}
\includegraphics[width = 3.5in]{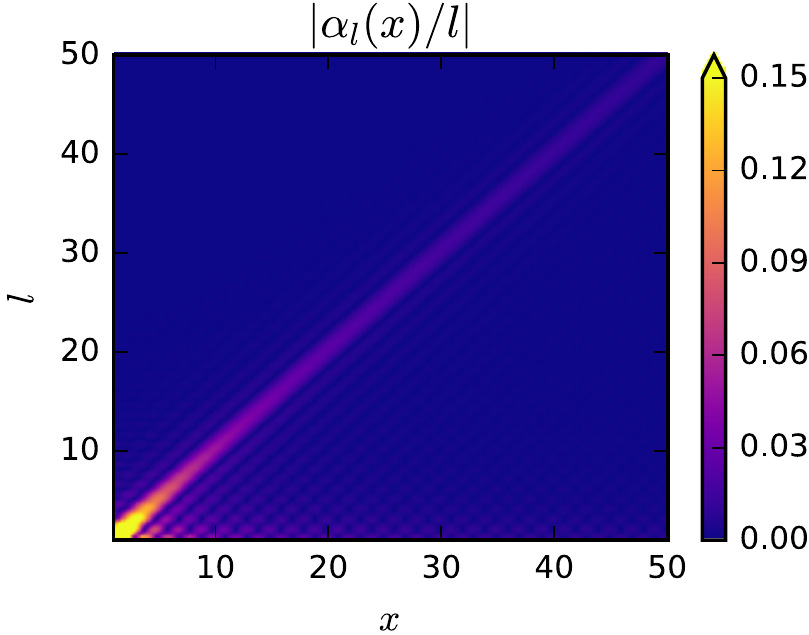}
\caption{Harmonic coefficients of $e^{ixF(\Omega t)}$ in Eqn. \ref{eq:alphaCoeff} for the square wave which control the periodic time-dependence in the rotating frame. The quick decay away from the peak at $l=x$ allows for a controlled $J/U$ expansion. }
\label{fig:alpha}
\end{figure}

Decomposing the rotated Hamiltonian into harmonics, $\tilde{H} = \sum_{l}e^{il\Omega t}\tilde{H}^{(l)}$, we obtain the relation $\tilde{H}^{(l)} = \sum_{m=0,\pm1}\tilde{H}_{m}\alpha_{l}(m\kappa)$ where $e^{im\kappa F(\Omega t)} = \sum_{l}e^{il\Omega t}\alpha_{l}(m\kappa)$ are the harmonic expansions of the time-dependent exponentials. For a square drive, we obtain
\begin{eqnarray}
\alpha_{l}(m\kappa) & = & \frac{i}{2\pi}\p{\frac{e^{-i\pi(l-m\kappa)}-1}{l-m\kappa}+\frac{1-e^{i\pi(l+m\kappa)}}{l+m\kappa}}
\label{eq:alphaCoeff}
\end{eqnarray}
and for a cosine drive, we obtain Bessel functions $\alpha_{l}(m\kappa) = \s J_{l}(m\kappa)$. Note that in the case of a square drive, the coefficients are peaked at $l=\pm m\kappa$ with power-law decay (see Figure \ref{fig:alpha}). This crucial property allows us to interpret the HFE as an effective (and approximate) $J/U$ expansion as we will see shortly. 

The general evolution operator for a Floquet system \cite{BukovPolkovnikovReview2015} has a periodic piece, $P(t,t_0)\equiv e^{-iK_{F}[t_{0}](t)}$, and a static piece $H_F[t_0]$, both of which depend on a choice of gauge $t_0$. The Hermitian operator $K_F[t_0](t)$ is known as the stroboscopic kick operator. Gauge transformations between choices of initial times are implemented with the micromotion operator $H_{F}[\tilde{t}_{0}] = P(\tilde{t}_{0},t_{0})H_{F}[t_{0}]P^{\dag}(\tilde{t}_{0},t_{0})$ where of course, by periodicity of $P$, initial times are only defined within a period.

Instead of choosing a single $t_0$, one might consider an alternative scenario where a symmetric gauge choice is selected such that no $t_0$ per se is favored. This particular gauge choice is useful if one wants to discuss a single Floquet Hamiltonian, which we will call the effective Hamiltonian, without the ambiguity of which initial time point was chosen. To this end, define a unitary transformation on the stroboscopic Floquet Hamiltonian $H_\mathrm{eff} = e^{iK_\mathrm{eff}(t_{0})}H_{F}[t_{0}]e^{-iK_\mathrm{eff}(t_{0})}$ such that the kick operators $K_\mathrm{eff}(t_0)$ are defined to rotate a given choice of stroboscopic Floquet Hamiltonian to the effective Hamiltonian. By using the gauge change formula for $H_F[t_0]$, one finds that $P(t,t_{0}) = e^{-iK_{F}[t_{0}](t)}=e^{-iK_\mathrm{eff}(t)}e^{iK_\mathrm{eff}(t_{0})}$, which immediately leads to the conclusion that the $H_\mathrm{eff}$ is indeed the static Hamiltonian obtained by rotating from the original frame with $e^{-iK_\mathrm{eff}(t)}$ instead of $P(t,t_0)$ which would yield $H_F[t_0]$. With these definitions, the general evolution has two representations
\begin{eqnarray}
U(t_{2},t_{1}) & = & P(t_{2},t_{0})e^{-iH_{F}[t_{0}](t_{2}-t_{1})}P^{\dag}(t_{1},t_{0})\nonumber\\
 & = & e^{-iK_{F}[t_{0}](t_{2})}e^{-iH_{F}[t_{0}](t_{2}-t_{1})}e^{iK_{F}[t_{0}](t_{1})}\nonumber\\
 & = & e^{-iK_\mathrm{eff}(t_{2})}e^{-iH_\mathrm{eff}(t_{2}-t_{1})}e^{iK_\mathrm{eff}(t_{1})}
\end{eqnarray}
where the kick and stroboscopic kick operators coincide if $K_\mathrm{eff}(t_0)=0$; this also means that the stroboscopic and effective Floquet Hamiltonian coincide. Quasienergy spectra are unaffected by kick operators since they are just a rotation of the Floquet Hamiltonian but measurement of observables requires one to take them into account.

In general, exact formulas for the effective Hamiltonians and kick operators are difficult to come by, so quite often one resorts to a high frequency expansion with $H_\mathrm{eff}$ encoding the gauge-symmetric Floquet Hamiltonian and $K_\mathrm{eff}$ encoding the explicit gauge change information. We will not rederive the results here and resort to quoting the series expansion for the effective Hamiltonian and kick operators up to second order from references \cite{GoldmanPRX,RahavPRA}. 
\begin{eqnarray}
H_{\mathrm{eff}} & = & H_{\mathrm{eff}}^{[0]}+H_{\mathrm{eff}}^{[1]}+H_{\mathrm{eff}}^{[2]}+... \label{eq:VVE-1}\\
K_{\mathrm{eff}} & = & K_{\mathrm{eff}}^{[0]}+K_{\mathrm{eff}}^{[1]}+K_{\mathrm{eff}}^{[2]}+...\label{eq:VVE-2}
\end{eqnarray}

In the main text, we have considered quasienergy states and spectra obtained from $U(T,0)$ and so kick operators used for numerical results are evaluated at $t=0$ (the particular gauge we have chosen for the stroboscopic Floquet Hamiltonian). We define $H_F^{[n]}$ as the $n$-th order approximation to the stroboscopic Floquet Hamiltonian obtained from the HFE.

\begin{widetext}

\begin{eqnarray}
H_{\mathrm{eff}}^{[0]} & = & \s H_{0}\nonumber\\
H_{\mathrm{eff}}^{[1]} & = & \frac{1}{\Omega}\sum_{j=1}^{\infty}\frac{1}{j}[\s V^{(j)},\s V^{(-j)}]\nonumber\\
H_{\mathrm{eff}}^{[2]} & = & \frac{1}{2\Omega^{2}}\sum_{j=1}^{\infty}\frac{1}{j^{2}}[[\s V^{(j)},\s H_{0}],\s V^{(-j)}]+\frac{1}{3\Omega^{2}}\sum_{j,l=1}^{\infty}\frac{1}{jl}([\s V^{(j)},[\s V^{(l)},\s V^{-(j+l)}]]-[\s V^{(j)},[\s V^{(-l)},\s V^{-(j-l)}]])+\mathrm{h.c.}\nonumber\\
K_{\mathrm{eff}}^{[0]}(t) & = & 0\nonumber\\
K_{\mathrm{eff}}^{[1]}(t) & = & \frac{1}{i\Omega}\sum_{j=1}^{\infty}\frac{1}{j}(\s V^{(j)}e^{ij\Omega t}-\s V^{(-j)}e^{-ij\Omega t}) \label{eq:VVE-full}\\
K_{\mathrm{eff}}^{[2]}(t) & = & \frac{1}{i\Omega^{2}}\sum_{j=1}^{\infty}\frac{1}{j^{2}}[\s V^{(j)},\s H_{0}]e^{ij\Omega t}+\frac{1}{2i\Omega^{2}}\sum_{j,l=1}^{\infty}\frac{1}{j(j+l)}[\s V^{(j)},\s V^{(l)}]e^{i(j+l)\Omega t}+\frac{1}{2i\Omega^{2}}\sum_{j\ne l=1}^{\infty}\frac{1}{j(j-l)}[\s V^{(j)},\s V^{(-l)}]e^{i(j-l)\Omega t}+\mathrm{h.c.}\nonumber
\end{eqnarray}

where $\s H_{0} = \tilde{H}^{(0)}$ and $\s V^{(j)} = (1-\delta_{j,0})\tilde{H}^{(j)} $. Inserting the harmonics of the rotated Hamiltonian, we obtain

\begin{eqnarray}
H_{\mathrm{eff}}^{[0]} & = & \tilde{H}_{0}+\sum_{m\ne0}\tilde{H}_{m}\alpha_{0}(m\kappa)\nonumber\\
H_{\mathrm{eff}}^{[1]} & = & \sum_{(m,m')\ne0}[\tilde{H}_{m},\tilde{H}_{m'}]\sum_{j=1}^{\infty}\frac{\alpha_{j}(m\kappa)\alpha_{-j}(m'\kappa)}{j\Omega}\nonumber\\
H_{\mathrm{eff}}^{[2]} & = & \sum_{(m,m')\ne0}[[\tilde{H}_{m},\tilde{H}_{0}],\tilde{H}_{m'}]\sum_{j=1}^{\infty}\frac{\alpha_{j}(m\kappa)\alpha_{-j}(m'\kappa)}{2\Omega^{2}j^{2}}\nonumber\\
 & + & \sum_{(m,m',m'')\ne0}[[\tilde{H}_{m},\tilde{H}_{m'}],\tilde{H}_{m''}]\\
 &  & \p{\sum_{j=1}^{\infty}\frac{\alpha_{j}(m\kappa)\alpha_{0}(m'\kappa)\alpha_{-j}(m''\kappa)}{2\Omega^{2}j^{2}}+\sum_{j,l=1}^{\infty}\frac{(1-\delta_{jl})\alpha_{-l}(m\kappa)\alpha_{-(j-l)}(m'\kappa)\alpha_{j}(m''\kappa)}{3\Omega^{2}jl}-\sum_{j,l=1}^{\infty}\frac{\alpha_{l}(m\kappa)\alpha_{-(j+l)}(m'\kappa)\alpha_{j}(m''\kappa)}{3\Omega^{2}jl}}+\mathrm{h.c.}\nonumber
\end{eqnarray}

\begin{eqnarray}
K_{\mathrm{eff}}^{[0]}(t) & = & 0\nonumber\\
K_{\mathrm{eff}}^{[1]}(t) & = & \sum_{m\ne0}\tilde{H}_{m}\sum_{j\ne0}\frac{\alpha_{j}(m\kappa)e^{ij\Omega t}}{ij\Omega}\nonumber\\
K_{\mathrm{eff}}^{[2]}(t) & = & \sum_{m\ne0}[\tilde{H}_{m},\tilde{H}_{0}]\sum_{j=1}^{\infty}\frac{\alpha_{j}(m\kappa)}{i\Omega^{2}j^{2}}e^{ij\Omega t}\\
 &  & +\sum_{(m,n)\ne0}[\tilde{H}_{m},\tilde{H}_{n}]\p{\sum_{j=1}^{\infty}\frac{\alpha_{j}(m\kappa)\alpha_{0}(n\kappa)}{i\Omega^{2}j^{2}}e^{ij\Omega t}+\sum_{j,l=1}^{\infty}\frac{\alpha_{j}(m\kappa)\alpha_{l}(n\kappa)}{2i\Omega^{2}j(j+l)}e^{i(j+l)\Omega t}+\sum_{j\ne l=1}^{\infty}\frac{\alpha_{j}(m\kappa)\alpha_{-l}(n\kappa)}{2i\Omega^{2}j(j-l)}e^{i(j-l)\Omega t}}+\mathrm{h.c.}\nonumber
\end{eqnarray}
\end{widetext}
where we have made use of the property $\alpha_{l}(0) = \delta_{l,0}$. Utilizing the peaking behavior of the $\alpha$ coefficients, we may understand the scaling of each term with $J/U$. We wish to compare the strength of each successive order of $H_\mathrm{eff}$ to the zeroth order term which scales as $J$. The first order, $H_\mathrm{eff}^{[1]}$, comes with a single commutator that yields two powers of $J$. Since each of the $\alpha$ coefficients are peaked when the subscript and arguments match (up to a sign), whenever the peaks of the two $\alpha$ coefficients overlap to give a nonzero contribution to the sum over $j$, they provide a scaling of $\kappa \Omega = U$ in the denominator, i.e.\mkadd{,} $j\Omega\rightarrow m\kappa\Omega=mU$. For a given system size $N$, the sum of the commutators provides some scaling with $N$ and so we denote the overall scaling of the first order term, relative to the zeroth order term, as $(J/U) f_1(N)$. Repeating the same arguments for the second order term gives three powers of $J$ and a denominator with two powers of $U$ for an overall relative scaling of $(J/U)^2 f_2(N)$. Each successive order gives one more power of $J$ due to an extra nested commutator
and another power of $U$ in the denominator due to replacement of some $j\Omega$
term with $U$. Therefore, up to errors introduced by the power law decay of the $\alpha$
coefficients, we have constructed an approximate $J/U$ expansion from the HFE.

\begin{figure*}

\includegraphics[width= \textwidth]{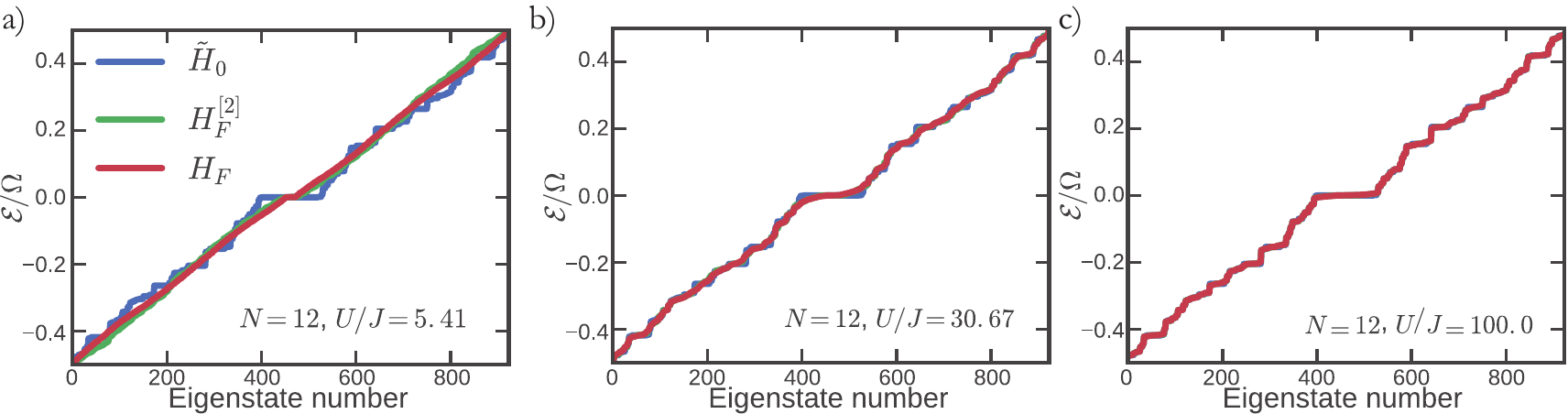}
\caption{
Quasienergies of the exact $H_{F}$ and the effective Hamiltonians
$\tilde{H}_{0}$ and $H_F^{[2]}$ for three values of $U/J$.
At large $U/J$, the spectra match, while away from this limit it
is clear that $H_F^{[2]}$ is a good approximation over
some region before the breakdown of the high frequency expansion.}
\label{fig:H_eff_spec}
\end{figure*}

The convergence properties and error bounds on such an expansion are largely unknown at this point in time. Two possibilities for such an expansion are that series is convergent or that it is asymptotic with an $n$ order expansion accurately describing the dynamics for some finite period of time, although recent work suggests that the latter is more likely\cite{Weinberg_Kolodrubetz_2017}. 

We delay further detailed analysis of this series and instead demonstrate the validity of our expansion by considering the large $U$ limit and comparing the exact Floquet spectrum to the spectrum of the effective Hamiltonian at various orders. Figure \ref{fig:H_eff_spec} shows the comparison of spectra between the exact Floquet Hamiltonian, zeroth (neglecting $H_{\pm1}$ terms - valid at large $U$), and second order effective Hamiltonian. Note that the first order term vanishes identically due to the symmetry $\alpha_j(m\kappa)=\alpha_{-j}(m\kappa)$ for the square wave. As expected, we see that for very large $U$, all the spectra match but as we decrease $U$, the zeroth order term deviates first before the second term which eventually also breaks down.

\section{Additional Evidence for Integrability and its Breaking \label{sec:Evidence} }

In this appendix, we provide additional evidence for integrability
of the effective high frequency model and for integrability-breaking
at finite $U$. Let us begin by showing how we numerically verify
integrability of $\tilde{H}_{0}$. As noted in the main text, $\tilde{H}_{0}$ is
a very unusual integrable model in the sense that multiple basis states
map to the same configuration in the language of the $\tilde{d}$ fermions,
leading to significant exact degeneracy. This is unlike simple free
models where no exact degeneracy exists, but rather a lack of level
repulsion allowing states to be close -- but not the same -- in energy.
Therefore, level statistics is not the ideal
test for integrability here.

\begin{figure}
\centering
\includegraphics[width = \columnwidth]{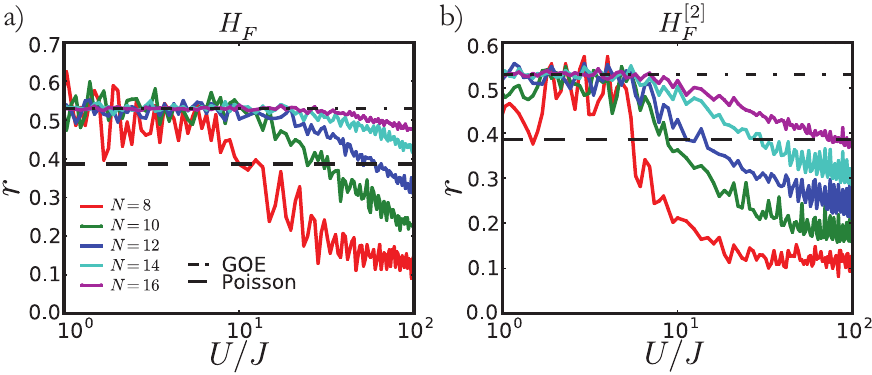}
\caption{\label{fig:level_stats} Level statistics of the exact $H_{F}$ and
the effective Hamiltonian $H_F^{[2]}$. For the static
Hamiltonian $H_F^{[2]}$, only the middle 50\% of the
spectrum is used to avoid noise from the often-anomalous high and
low energy tails. For small $U/J$, the level statistic is GOE indicating non-integrable behavior of the system. As $U/J$ increases, the level statistic breaks away from GOE indicating a different spectral structure due to near-integrability. Note that this crossover is system size dependent as seen clearly in a). In b), there is a much weaker system size dependence suggesting that the HFE, at second order, does not accurately capture the crossover from integrability to non-integrability.}
\end{figure}

Instead, we simply show that the spectrum may be reproduced by free
fermion numerics. The procedure to generate the spectrum of $\tilde{H}_{0}$
is as follows:
\begin{enumerate}
\item Iterate through basis elements of the original model.
\item For each basis element, map it to a representation in the $\tilde{d}$-basis.
\item In the $\tilde{d}$-basis representation, count the number of fermions and
the number of sites. For free one-dimensional fermions hopping on
such a lattice, calculate the spectrum.
\item Impose a degeneracy on the free fermion given by the number of original
basis elements that map to the same number of fermions and sites in
the $\tilde{d}$ representation.
\end{enumerate}
The spectrum obtained by this procedure is plotted as the $\tilde{H}_{0}$
data in Figure~\ref{fig:H_eff_spec}. For comparison, the spectra of
$H_{F}$ and $H_F^{[2]}$ are obtained through exact
diagonalization. The results clearly converge in the $U/J\to\infty$
limit, demonstrating the integrability of our model.
 
While level statistics is difficult for identifying the integrable
limit of our model, it remains the smoking gun for seeing the breaking
of integrability. In Figure~\ref{fig:level_stats} we show the level
statistic $r=\mathrm{min}(\Delta E_{n},\Delta E_{n+1})/\mathrm{max}(\Delta E_{n},\Delta E_{n+1})$
for the exact and effective Floquet Hamiltonian, where $\Delta E_{n}\equiv E_{n}-E_{n-1}$
is the (quasi)energy difference between Floquet eigenstates $n$ and
$n-1$. It has been well-studied that this object crosses over from
$r\approx0.386$ (Poisson statistics) to $r\approx0.53$ (Gaussian
orthogonal ensemble, a.k.a. GOE) as the system crosses from integrable
to non-integrable \cite{Oganesyan_Huse_2007,Ponte2014b}.
The non-integrable plateau is clearly seen for both $H_{F}$ and $H_F^{[2]}$,
indicating that both obey the eigenstate thermalization hypothesis
for a finite range of $U$. We also see that, due to the unusual nature
of the integrable model, the Poisson limit is not reached at very
large $U$. Similar to crossover behavior of $\Sigma$ in the main
text, the level statistics show a system size dependent crossover
for both $H_{F}$ and $H_F^{[2]}$ (albeit much weaker for $H_F^{[2]}$), consistent with our belief that the both $H_F$ and $H_F^{[2]}$ will thermalize for infinitesimal finite $J/U$ in the thermodynamic limit.

\begin{figure*}

\includegraphics[width= \textwidth]{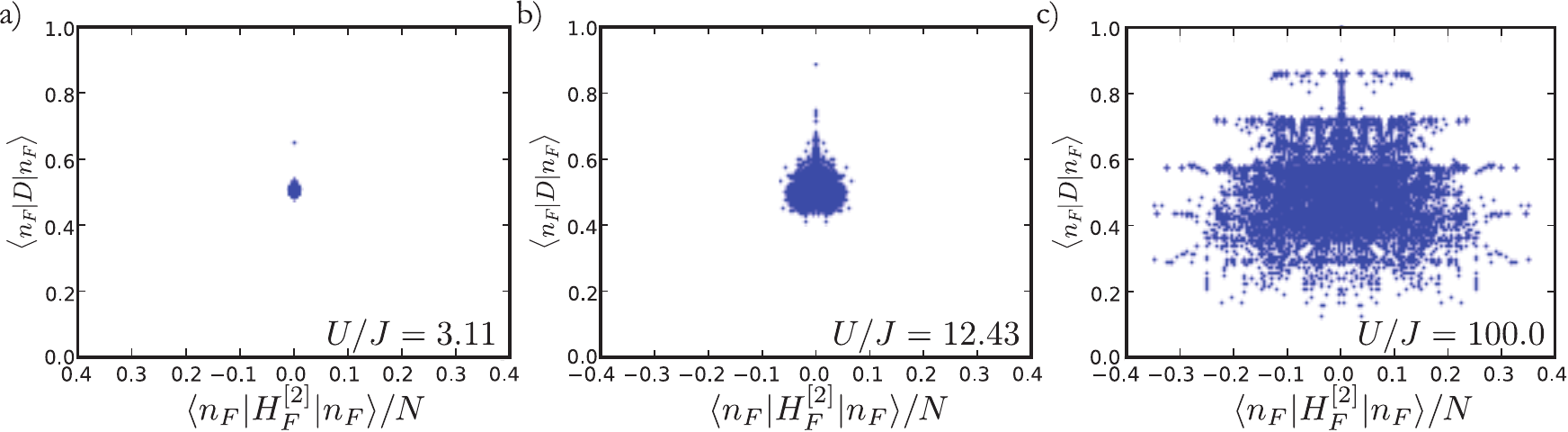}
\caption{
Expectation values of the doublon density $D$ and effective HFE Floquet
Hamiltonian $H_{F}^{[2]}\equiv e^{-iK_{\mathrm{eff}}^{[2]}(0)}H_{\mathrm{eff}}^{[2]}e^{iK_{\mathrm{eff}}^{[2]}(0)}$
in exact Floquet eigenstates of $H_{F}$. 
\label{fig:eev}}
\end{figure*}

Furthermore, we provide additional evidence that the crossover to thermalization
of $H_\mathrm{eff}$ at finite $U$ is not governed by the breaking of integrability
in $H_F^{[2]}$, but rather by a direct breakdown of
the high frequency expansion. In Figure~\ref{fig:eev} we plot eigenstate
expectation values of two observables: the doublon density $D$ and
the HFE Hamiltonian $H_F^{[2]}$. As a local observable, we expect
$D$ to satisfy the Floquet-ETH for $U$
beyond the thermalization crossover, meaning that eigenstate expectation
values of $D$ should be independent of quasienergy and with fluctuations exponentially
suppressed in system size. This is consistent with the data shown,
as $D$ compresses into a narrower region as $U$ is decreased, 
approaching the single value $D=1/2$ in the thermodynamic
limit. On the other hand, if $H_F^{[2]}$ were a good description
of the system in this non-integrable region, we would expect that
$H_{F}^{[2]}$ would become nearly conserved, implying
that its expectation value would be extensively spread over eigenstates.
Instead, we see that $H_F^{[2]}$ behaves exactly as $D$, approaching
a single point in the non-integrable limit. This implies that $H_F^{[2]}$
is not a conserved quantity in the system, and thus behaves exactly
the same as other non-conserved quantities such as $D$ that satisfy
the Floquet-ETH.

\begin{figure}
\includegraphics[width= 3.5in]{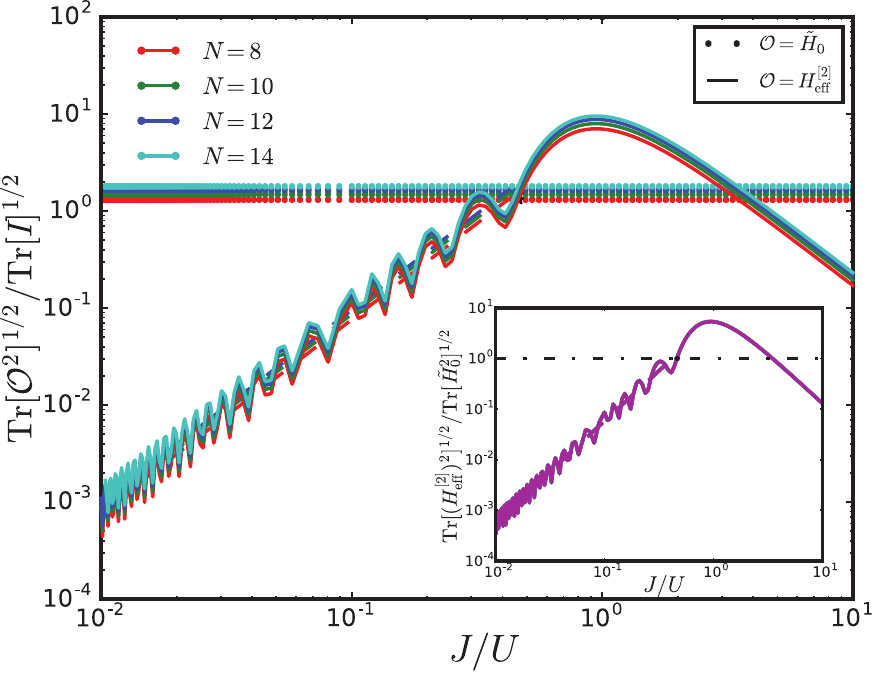}
\caption{\label{fig:tracenorm}
Frobenius norm of the integrable model $\tilde{H}_0$ and of the second order term $H_\mathrm{eff}^{[2]}$ in the HFE normalized to the trace norm of the identity for each system size. As discussed in Appendix \ref{sec:Expansion}, the trace norm has $(J/U)^2$ behavior indicated by the dashed line. At $J/U \sim 0.5$, the second order term is relatively larger than the integrable part. Both zeroth order and second order terms have the same system size dependence as seen by observing the relative trace norm in the inset. This fact immediately rules out the possibility of that the breakdown of the HFE as an operator expansion is responsible for thermalization as discussed in section \ref{sec:Integrability and its breakdown} and appendix \ref{sec:Evidence}, at least at second order. }
\end{figure}

Finally, let us see that the breakdown of the HFE is due to the resonances discussed in the main text and not directly due to a breakdown of the operator series for $H_\mathrm{eff}$. For finite size systems, the expansion in \eqref{eq:VVE-1} should have a well-defined convergence radius in the space of finite-dimensional matrices. We can look for the breakdown of this series by directly comparing the size of the leading correction, $H_\mathrm{eff}^{[2]}$, to the zeroth order term $\tilde{H}_0$.\footnote{We have numerically confirmed that the $\tilde{H}_{\pm 1}$ corrections to $H_\mathrm{eff}^{[0]}$ play a sub-leading role in all of the analyses in this paper.} This is achieved in Figure~\ref{fig:tracenorm} by comparing their Frobenius norms. These norms collapse amazingly well as a function of system size, such that we can immediately conclude that $H_\mathrm{eff}^{[2]}$ becomes of order $\tilde{H}_0$ at fixed ratio $J / U \sim 0.5$ independent of system size. Thus we conclude that, at least to second order, there is a finite system size independent radius of convergence for the HFE, which is clearly in conflict with the breakdown of integrability in the exact $H_F$. This provides additional evidence that the breakdown of integrability is due to non-perturbative effects for our choice of parameters, though it is possible that the direct breakdown of the HFE series expansion will be the leading effect for other models or values of the parameters.

\section{Alternative mapping}
In this section, we present an alternative, but equivalent, mapping that demonstrates the integrability of $\tilde{H}_0$. First, define a defect as a single site that is surrounded by sites of the opposite kind. A domain wall will be two sites that are part of a sequence of occupation longer than 1. For instance: 0(01)1 is a domain wall. Now let us define the contracted lattice as a lattice where in each site we can have a hole, 0, a particle, 1, a domain wall (01) or (10) which we will call W, and a defect which is either (10) in a 111 domain or a (01) in a 000 domain which we call D. For a particular collection of these objects, we can have a lattice exemplified as follows:
\[
000000111110000 \rightarrow 00000W111W000
\]
and with defects:
\[
001000111010000 \rightarrow 0D00W1DW000
\]
A defect can move freely as long as there is no domain wall on the site it ends on. If there is a domain wall, then they switch positions:
\[
1101000=1DW00  \rightarrow 1100100=1WD00
\]

This means that we can write the hamiltonian as follows. We define $d$'s as annihilation operators for the defects, and $b$'s as annihilation operators for the domain walls. 
\begin{equation}
H=d^{\dagger}_{i+1}d_i\left((1-n^W_{i+1})+b_{i}^{\dagger}b_{i+1}\right)+h.c.
\end{equation}
It is also clear that defects are hard-core bosons. This almost looks like free hard-core bosons except for the shift in location of the domain wall. This can be taken into account by an appropriate string operator which we now construct. First, consider the following unitary:
\begin{equation}
U_{j}=\left[1+(-1)^{n^W_i+n^W_{i+1}}+2\left(b^{\dagger}_i b_{i+1}+b^{\dagger}_{i+1} b_{i}\right)\right]
\end{equation}
If there is no domain wall on site $i+1$, then this will shift a domain wall at site $i$ to site $i+1$. We envision the chain as terminating by some domain, with no walls, so if we have a string starting operation from the farthest point, and counting to the left, this will shift all domain walls one step to the right. Alternatively, if we start from the location to the left of a domain wall, and multiply the unitaries into a string, we will shift all domain walls to the left. So if we define:
\begin{equation}
\tilde{d}_i=\left[\prod\limits_{j=N}^{i} U_j\right] d_i
\end{equation}
we immediately get:
\begin{equation}
H=\tilde{d}^{\dagger}_{i+1}\tilde{d}_i+h.c.
\end{equation}
which is the integrable model. We may further consider the creation of defects at finite $U$. This can happen only in the vicinity of a domain wall. To construct the operator we consider a $U$ energy step: 
\[
\begin{array}{c}
111W000=111(10)000\\
\downarrow \\
11101000=11(10)_D(10)_W00=11DW00
\end{array}
\]
with the upshot that now a site is missing. A chain with 7 effective sites, now only has 6 due to the contraction that the mapping of the defect implies. We can describe this process as originating from: 
\begin{equation}
H_U=   d^{\dagger}_{i-1} W_i b^{\dagger}_{i+1} b_i +h.c.
\end{equation}
where the $W_i$ is a ``warp" operator which moves everything to the left and cancels site $i$. One can write it in terms of string operator for both domain walls and defects. There is an implicit gauge choice in the above in the sense that domain walls created defects to their left, regardless of their nature.

\bibliographystyle{apsrev}
\bibliography{FloqMottRefs2}

\end{document}